\documentclass[aps,prx,amsmath,amssymb,reprint,author-year,author-numerical,floatfix,superscriptaddress]{revtex4-2}
\usepackage{graphicx}
\graphicspath{{./images/}}
\usepackage{dcolumn}
\usepackage{bm}
\usepackage{mathtools}
\usepackage[caption=false]{subfig}
\usepackage{amssymb}
\expandafter\let\csname equation*\endcsname\relax
\expandafter\let\csname endequation*\endcsname\relax
\usepackage{amsmath}
\usepackage{amsbsy}
\usepackage{booktabs}
\usepackage{verbatim}
\usepackage{array}
\newcolumntype{P}[1]{>{\centering\arraybackslash}p{#1}}
\usepackage[colorlinks=true,citecolor=blue]{hyperref}

\AtBeginDocument{\renewcommand{\natexlab}[1]{#1}}%

\begin{document}


\title{A first-principles approach to closing the ``10-100 eV gap'' for charge-carrier thermalization in semiconductors}


\author{Dallin O. Nielsen}
\address{The University of Texas at Dallas, Department of Materials Science and Engineering, 800 W. Campbell Rd., Richardson, Texas 75080, USA}

\author{Chris G. Van de Walle}
\address{University of California, Santa Barbara, Materials Department, 2510 Engineering II, Santa Barbara, CA 93106-5050, USA}

\author{Sokrates T. Pantelides}
\address{Vanderbilt University, Department of Physics and Astronomy, Vanderbilt University, Nashville, TN 37235-1824, USA}

\author{Ronald D. Schrimpf}
\address{Vanderbilt University, Department of Electrical and Computer Engineering, VU Station B 351824, 2301 Vanderbilt Place, Nashville, TN 37235-1824, USA}

\author{Daniel M. Fleetwood}
\address{Vanderbilt University, Department of Electrical and Computer Engineering, VU Station B 351824, 2301 Vanderbilt Place, Nashville, TN 37235-1824, USA}

\author{Massimo V. Fischetti}
\address{The University of Texas at Dallas, Department of Materials Science and Engineering, 800 W. Campbell Rd., Richardson, Texas 75080, USA}


\date{\today}

\begin{abstract}
Since the 1960s and the first observations of radiation-induced disruption of electronic devices in space, the study of the effects of ionizing radiation on electronics has grown into an extensive field of its own. The present work is concerned with studying accurately the energy-loss processes that control the thermalization of hot carriers (electrons, holes, and/or electron-hole pairs) that are generated by high-energy radiation in wurtzite GaN, using an {\it ab initio} approach. Current physical models of the nuclear/particle physics community cover thermalization in the high-energy range (kinetic energies exceeding $\sim$~100~eV), and the electronic-device community has studied extensively carrier transport in the low-energy range (below $\sim$~10~eV).  However, the processes that control the energy losses and thermalization of electrons and holes in the intermediate energy range of about 10-100 eV (which we define as the “10-100 eV gap”) are poorly known. The aim of this research is to close this gap. To this end, we utilize density functional theory (DFT) to obtain the band structure and dielectric function of GaN for energies up to about 100~eV. We also calculate charge-carrier scattering rates for the major charge-carrier interactions (phonon scattering, impact ionization, and plasmon emission), using the DFT results and first-order perturbation theory (Fermi’s Golden Rule/first Born approximation). With this information, we study the thermalization of electrons starting at 100~eV using the Monte Carlo method to solve the semiclassical Boltzmann transport equation. Full thermalization of electrons and holes is complete within $\sim$~1 and 0.5~ps, respectively. Hot electrons dissipate about 90\% of their initial kinetic energy to the electron-hole gas (90 eV) during the first $\sim$~0.1~fs, due to rapid plasmon emission and impact ionization at high energies. The remaining energy is lost more slowly as phonon emission dominates at lower energies (below $\sim$~10~eV). During the thermalization, hot electrons generate pairs with an average energy of 
$\sim$~8.9~eV/pair (11-12 pairs per hot electron). Additionally, during the thermalization, the maximum electron displacement from its original position is found to be on the order of 100 nm.
\end{abstract}


\maketitle

\section{Introduction}
\label{sec:intro}
In the study of ionizing radiation effects (including total ionizing dose (TID) and single-event effects (SEE)) in electronic devices and that of radiation detection ({\it i.e.}, scintillators and semiconductor detectors), many computational tools have been developed to simulate particle transport and also the resulting possible material damage \cite{Reed15,Agostinelli03,Allison06,Biersack80}. These codes, developed by the nuclear/particle physics community, typically employ the binary collision and free-electron approximations. The primary assumption of the binary collision approximation is that the energetic projectile (an ion, for example) interacts via a series of independent two-body interactions with atoms in the material. As the energy decreases, this approximation begins to break down, as simultaneous interactions with multiple atoms occur. To handle this issue, these codes may consider nearly simultaneous interactions, allowing their use to be extended down below the keV range. At still lower energies ($\sim$~100~eV), the accuracy of the simulations comes into question as their results do not agree with experimental measurements, which yield deviations from monotonic response. For example, in Ref.~\cite{Dozier81}, a monotonic decrease in the charge yield is measured with decreasing energy until energies reach the order of hundreds of eV. At this point, the charge yield apparently increases as the energy decreases from $\sim$~100~eV to $\sim$~70~eV. It is not clear what causes this phenomenon, but it is reasonable to suspect that electronic band structure effects may be partly to blame, and that inaccuracies of the simulations are associated with the use of the free-electron approximation. In addition, inaccuracies are likely caused by an inaccurate treatment of charge-carrier interactions.
 
The primary concern of this work is to study accurately the thermalization of hot carriers (electrons and/or electron-hole pairs) generated by high-energy radiation in semiconductors at energies below this threshold of $\sim$~100~eV. Akkerman {\it et al.} have developed a code that employs the binary collision and free-electron approximations to simulate electron transport down to $\sim$~20~eV (suggesting that the code can be extended down to 5~eV in some cases) \cite{Akkerman09}. Their approach, however, does not take the band structure and carrier-phonon scattering into account, which likely are critical for accurate simulation (especially for energies below $\sim$~10~eV)~\cite{fang_2019}.

To account for band structure effects, one may look to the electronic device community, which has done extensive work in developing codes to simulate carrier transport in the “low-energy” region (kinetic energies below $\sim$~10~eV) \cite{MVF88,Fang19,Ghosh17,Bertazzi09,reaz_2021}. Such simulations employ the full band structure in solving the semi-classical Boltzmann transport equation via the Monte Carlo (MC) method. 

Between this low-energy region and what we call the “high-energy” region (kinetic energies exceeding $\sim$~100~eV), there is an “intermediate-energy” range of about 10-100 eV, where the processes that control the energy losses and thermalization of electrons and holes are poorly known. We define this region as the “10-100~eV gap”. While only scant information on this topic is available, a major study by Pines showed that losses to plasmons are the dominant process \cite{Pines_1956}. We aim to study the scattering processes in this regime rigorously from first principles. In this effort, we include plasmon losses as well as impact ionization and phonon scattering in our model. To close this ``10-100~eV gap'', we utilize the full-band approach and build up to $\sim$~100~eV.

\begin{figure}[!]
\includegraphics[height=3.23in, width=2.43in,clip]{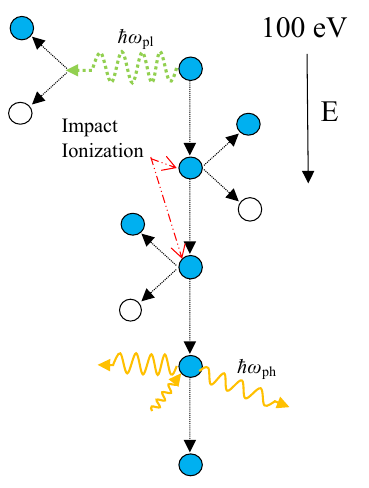}
\caption{\label{fig:therm_proc} A diagram illustrating the processes that occur during the thermalization of the hot carriers in our simulation. Solid (blue) and open circles represent electrons and holes, respectively, while the 'wiggly' dashed and solid lines (green and yellow) represent plasmons ($\hbar\omega_{\rm pl}$) and phonons ($\hbar\omega_{\rm ph}$), respectively.}\end{figure}

The energy-loss processes in the ``10-100 eV gap'' have been studied theoretically in the past: Rothwarf~\cite{rothwarf_1973} and Kingsley and Ludwig~\cite{kingsley_1970} in phosphors, Alig {\it et al.} in semiconductors~\cite{alig_1980}, Ausman and McLean  in SiO$_2$~\cite{ausman_1975}. These earlier studies followed Pines~\cite{Pines1956a} in assuming that the main energy losses were due to plasmons. They considered also their decay into electron-hole pairs (EHPs), generation of EHPs via impact ionization processes, and final thermalization of the carriers via phonon scattering (see Fig.~\ref{fig:therm_proc}). However, these calculations were based on simplifying assumptions, such as the use of semi-empirical matrix elements and/or scattering rates and, most important, the free-electron model.  Here we take a similar approach but use {\it ab initio} methods to compute both the band structure and the energy-loss and scattering rates. Indeed, it may be argued that the cutoff energy used in density functional theory (DFT), typically as large as 80~Ry, may be an excessive overestimation of the energy above which electrons can safely be taken as free. However, for electron kinetic energies lower than about 100~eV the ionic (pseudo)potentials represent a fraction of the total energy so large, 10\% or more, that it cannot be ignored and the free-electron model cannot be expected to be accurate.

In addition to these early plasmon-based studies, with the prediction of single-event upset in 1962 \cite{Wallmark62}, and the subsequent discoveries of other types of SEE \cite{Kolasinski79, Wyatt79,Guenzer79,soliman_1983,shiono_1986} and TID, the study of ionizing radiation effects in Si-based devices has received much attention \cite{Kolasinski79,Wyatt79,Guenzer79,shaneyfelt_1991,oldham_2003,schwank_2008}. Due to a relatively recent (within the past couple decades) surge in interest in wide band gap semiconductors \cite{Bose13,Franquelo08,Reynolds12}, however, much work has been reported on ionizing radiation effects in materials such as SiC \cite{Akturk17,Witulski18,Ball20,Zhou19}, diamond, $\beta$-Ga$_2$O$_3$~\cite{ma_2023}, and GaN \cite{Fleetwood22,ives_2015,jiang_2017}. As the use of wide band gap materials expands, there is a growing demand for implementation in space. For this work, we have chosen to focus on wurtzite GaN. In principle, however, the methods described in this paper can be applied to any semiconductor of interest.

GaN has a significantly larger band gap than Si (3.4 eV compared to 1.1 eV). The larger band gap gives it a much higher breakdown field (3.3 MV/cm compared to 0.3 MV/cm), which makes GaN a good candidate for high power electronics and extreme device scaling. GaN has a comparable, if not somewhat higher, electron mobility ($\sim$~1300-2000~cm$^2$V$^{-1}$cm$^{-1}$ compared to 1440-1500~cm$^2$V$^{-1}$cm$^{-1}$) due to its relatively small effective mass of 0.2~$m_{\rm e}$ (the free electron mass). The electron drift velocity in GaN reaches a peak of $\sim 2.5$-$3.0 \times 10^{7}$~cm/s, while in Si, it saturates at $\sim 1.0 \times 10^{7}$~cm/s. These characteristics have made GaN-based devices an increasingly important technology over the past couple decades. 

As mentioned above, we use an {\it ab initio} approach (DFT) to obtain the electronic band structure of wurtzite GaN for bands reaching energies above 100~eV. Relevant charge-carrier scattering rates are calculated for the major charge-carrier interactions (phonon scattering, impact ionization, and plasmon emission), using the DFT results and first-order perturbation theory (Fermi’s Golden Rule/first Born approximation). With this information, the thermalization of electrons is simulated starting at 100~eV in a full-band MC (FBMC) code. A number of results are extracted, including the average carrier energy, energy distribution and position as a function of time, the average energy per generated pair, etc. 

It is important to note that here we restrict our attention to low-dose irradiation at a low-dose-rate and, so, a low density of carriers. This restriction allows us to assume that the number of generated plasmons is small enough to leave their distribution at thermal equilibrium, so that we can ignore plasmon absorption. Similarly, the low density of carriers permits us to ignore also short-range carrier-carrier scattering.

\section{First-Principles Calculations}
\label{sec:method}
\subsection{Electronic Band Structure}
\label{sec:method.abinitio}
For this work, we use the DFT package Quantum ESPRESSO (QE)~\cite{giannozzi_2009} for all {\it ab initio} calculations. Norm-conserving pseudopotentials \cite{Schlipf15} with PBE \cite{Perdew96} exchange-correlation (XC) functionals are employed. The unit cell of wurtzite GaN is hexagonal with a space group of P6$_3$mc. We have utilized the experimentally measured lattice constants of $a_0=3.215\,{\rm \AA}$ and $c_0=5.241\,{\rm \AA}$ \cite{Lagerstedt79}. For the self-consistent calculation, we use a plane-wave cutoff energy of 120~Ry and a uniform $8\times8\times8$ Monkhorst-Pack grid of {\bf k} points.

The FBMC simulation requires knowledge of the electronic band structure everywhere in the first Brillouin zone (BZ). We calculate the band structure for a total of 150 bands (up to $\sim$~100~eV above the conduction band edge) for 3234 {\bf k} points in the irreducible wedge of the BZ. These points are obtained by applying the 24 symmetry operations of wurtzite GaN to a uniform $40\times40\times40$ Monkhorst-Pack grid. Of the 150 bands, the first 18 are valence bands, representing the states of the 18 valence electrons of GaN (10 {\it d}- and 8 {\it sp}-electrons). The remaining 132 are conduction bands. 

We note that according to x-ray photoemission spectroscopy measurements \cite{martin_1994}, below the $d$ states, the next valence state of GaN is the Ga $3p_{3/2}$ state at $-102.6\pm0.1$~eV (with respect to the valence band edge). To meet the requirement for energy conservation, a charge carrier would need to possess a minimum energy of 105.0~eV to excite an electron from this core state into the conduction bands. Taking momentum conservation into account as well, the threshold energy would be significantly higher. Thus, we may safely assume that core electrons do not play a significant role in the thermalization of charge carriers in the ``10-100 eV gap'' in GaN.
\begin{figure}[!]
\includegraphics[height=4in, width=3.25in]{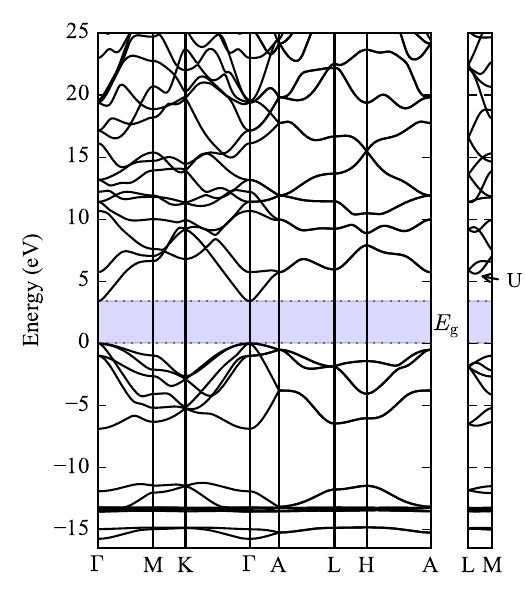}
\caption{\label{fig:band_structure}The band structure of wurtzite GaN along several symmetry lines of the BZ is shown. The primary band gap is indicated by the shaded region labeled $E_{\rm g}$. The symmetry line L-M is included for the reader to see the U valley (the lowest-energy valley above the gap, along this line).}
\end{figure}

Figure~\ref{fig:band_structure} shows the band structure calculated along several high-symmetry lines of the BZ. The primary band gap is indicated by the shaded region labeled $E_{\rm g}$. Below this gap, all 18 valence bands are shown. These can be divided into three major groups. The group of the six highest-energy valence bands ($-7$ to 0~eV) represents {\it sp} states. Below these, there is a gap followed by a group of two more {\it sp} bands, which intersects a dense set of eight {\it d} bands. The final two {\it d} bands appear below these at around $-$15~eV. We note the detailed positions of these bands as they may vary with the pseudopotential and XC used.
\begin{table*}[!]
\caption{\label{tab:table1}Calculated vertical transition energies (in eV) across the energy gap at certain symmetry points of wurtzite GaN with comparisons to available theoretical and experimental data from~\cite{Lambrecht95,Brockt00,Logothetidis94}.}
\vspace*{0.35cm}
\begin{ruledtabular}
\begin{tabular}{lccccccc}
&U$_{\rm v_1-c_1}$ &M$_{\rm v_1-c_2}$ &L$_{\rm v_1-c_1}$ &K$_{\rm v_1-c_1}$ &K$_{\rm v_2-c_1}$ &A$_{\rm v_1-c_3}$ &L$_{\rm v_3-c_3}$ \\ 
\colrule
This work&7.18&8.00&7.80&9.42&9.68&10.49&11.10\\
DFT-LDA~\cite{Lambrecht95}&6.87&7.65&7.64&9.57&9.68&10.53&11.05\\
Expt.~\cite{Lambrecht95}&6.9&8.0&8.0&9.3&9.3&10.5\text{-}11.5&10.5\text{-}11.5\\
Expt.~\cite{Brockt00}&7.1&8.1&8.1&9.2&9.2&\text{-}&\text{-}\\
Expt.~\cite{Logothetidis94}&7.0&7.9&7.9&9.0&9.0&\text{-}&\text{-}\\
\end{tabular}
\end{ruledtabular}
\end{table*}
\begin{figure}[!]
\includegraphics[height=3in, width=3in]{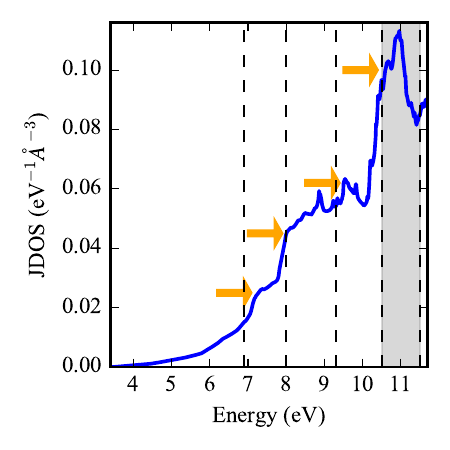}
\caption{\label{fig:jdos}Calculated JDOS of GaN from DFT, as a function of energy. The dashed lines indicate the locations of vertical energy transitions across the gap measured by Lambrecht {\it et al.}~\cite{Lambrecht95}. Roughly corresponding peaks and shoulders in the JDOS are indicated with arrows (orange).}
\end{figure}

The DFT-calculated primary gap is 1.78~eV, which is slightly more than half of the measured value of 3.4 eV. This issue of underestimating the gap is a well-known problem associated with DFT. Many ``solutions'' have been proposed, including hybrid XC functionals, $G_{0}W_{0}$, and others. These have been highly successful in calculating the proper band gap, but their effects on transport properties are inconsistent~\cite{Nielsen21}. Here we simply employ the ``scissors'' operator, shifting the conduction bands up by 1.62~eV to correct the gap. Additionally, using a curve-fitting technique, we calculate an electron effective mass of $\sim0.17~m_{\rm e}$ in the $\Gamma$ valley, which is somewhat smaller than the experimental value of $(0.20 \pm 0.005)~m_{\rm e}$~\cite{Drechsler95}. It is, however, similar to the DFT result of 0.18~$m_{\rm e}$ from Ref.~\cite{suzuki_1996}.

To assess the validity of the calculated band structure, we obtain the energies of direct transitions across the gap at certain symmetry points by calculating the joint density of states (JDOS):
\begin{equation}
JDOS(E)=\frac{2}{\Omega_{\rm c}} \sum_{nm{\bf k}}\delta[E-(E_{n}({\bf k})-E_{m}({\bf k}))], 
\label{eq:JDOS}
\end{equation}
where $\Omega_{\rm c}$ is the unit cell volume, $n$ is the conduction band index and $m$ in the valence band index. 
``Peaks'' and ``shoulders'' of the JDOS occur for vertical transitions at BZ symmetry points. Similar peaks and shoulders occur in measured absorption/reflectivity spectra and dielectric function data, allowing one to make a direct comparison. 

To evaluate the summation over the delta function in Eq.~(\ref{eq:JDOS}), we use Bl{\"o}chl's tetrahedron method \cite{Blochl94}. It involves a discretization of the first BZ along the reciprocal lattice vectors (using the same uniform grid from above), filling it with ``cube'' elements. Each element is then divided into six tetrahedra. We then scan the BZ for all cubes which obey the energy-conservation requirement in Eq.~(\ref{eq:JDOS}). For all such cubes, the density of states (DOS) is evaluated by summing the contributions from each tetrahedron.

In Ref.~\cite{Lambrecht95}, Lambrecht {\it et al.} measured reflectivity curves for wurtzite GaN. The resulting data yield several major peaks for which they identified corresponding transitions across the gap. A few of these major peaks in the reflectivity plot occur at 6.9~eV, 8.0~eV, 9.3~eV, and between 10.5 and 11.5~eV. In Fig.~\ref{fig:jdos}, we show the calculated JDOS; the above experimental energies and energy range have been marked with dashed lines. Arrows indicate characteristic peaks and shoulders. Qualitatively, we observe good agreement with Ref.~\cite{Lambrecht95}, especially for the peaks at $\sim$~8~eV and $\sim$~11~eV.
\begin{figure}[!]
\includegraphics[height=4in, width=3.5in]{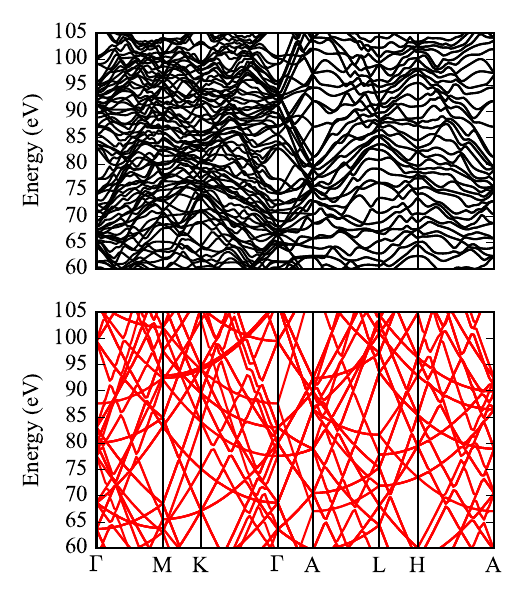}
\caption{\label{fig:free_bands}Top: Band structure from DFT at kinetic energies near 100 eV. Bottom: The free-electron dispersion within the empty lattice approximation at kinetic energies near 100 eV.}
\end{figure}
\begin{table*}[!]
\caption{\label{tab:table2}Calculated phonon energies in meV for several phonons at the $\Gamma$ point. Each phonon is given a label (E$_2$, B$_1$, etc.) that can be found in Fig.~\ref{fig:ph_disp}, where the associated location is clearly indicated.}
\vspace*{0.35cm} 
\begin{ruledtabular}
\begin{tabular}{lcccccccc}
 &
E$_{2}$ (low)&
B$_{1}$ (low)&
A$_{1}$ (TO)&
E$_{1}$ (TO)&
E$_{2}$ (high)&
B$_{1}$ (high)&
A$_{1}$ (LO)&
E$_{1}$ (LO)\\
\colrule
This work&16.89&39.94&63.35&65.39&66.31&82.34&86.56&86.73\\
Theory \cite{bungaro_2000}&17.11&41.41&68.19&70.92&71.17&85.55&90.88&91.37\\
Theory~\cite{karch_1998}&17.73&41.78&67.07&70.42&71.79&89.27&92.74&93.85\\
Theory~\cite{gorczyca_1995}&18.60&40.91&66.58&68.81&69.18&83.94&&\\
Expt. \cite{Ruf01}&&40.79&&&&85.79&90.39&\\
Expt. \cite{Azuhata95}&17.85&&66.08&69.55&70.55&&91.13&92.12\\
\end{tabular}
\end{ruledtabular}
\end{table*}

We list in Table~\ref{tab:table1} the vertical transitions identified by Lambrecht {\it et al.} as well as those measured in Ref.~\cite{Brockt00} (via electron energy-loss spectroscopy) and Ref.~\cite{Logothetidis94} (via synchrotron ellipsometry). In addition, we include the DFT-LDA work by Lambrecht {\it et al.}~\cite{Brockt00}. In the table, c$_{i}$ and v$_{i}$ refer to conduction bands and valence bands, respectively, with $i$ representing the index of the band. For the valence band indices, $i=1$ corresponds to the highest valence band, and the index increases for deeper bands. For the conduction band indices, $i=1$ corresponds to the lowest conduction band and increases for higher bands. The point U refers to the satellite valley between L and M. 

We note that for the above experimental work, the peaks associated with transitions have relatively large widths, leading to uncertainty in the true transition energies. This uncertainty leads, for example, to an apparent degeneracy at K$_{\rm v_1}$ and K$_{\rm v_2}$ in the experimental results, which does not agree with either our or the LDA calculations. We can therefore say only that the measured peaks give approximate energies of the transitions, and that the DFT data from this work and Ref.~\cite{Lambrecht95} are all in reasonably good agreement with these energies. Indeed, the largest discrepancy, occurring between our work and Ref.~\cite{Logothetidis94} for K$_{\rm v_2{\text -}c_1}$, of 0.68 eV is not much larger than the discrepancies between the experimental results themselves (as large as 0.3 eV). Additionally, our calculated energy for this transition (9.68 eV) easily falls within the widths of the corresponding peaks (9.3 and 9.2 eV, respectively) in the plots of Refs.~\cite{Lambrecht95,Brockt00}. Thus, we conclude that the use of the pseudopotential-XC functional combination of ONCV+PBE to calculate the band structure is justified. 

Lastly, we return briefly to the discussion on the use of DFT bands instead of free-electron bands. In Fig.~\ref{fig:free_bands}, we plot both the DFT (conduction) bands (top frame) and the free-electron dispersion within the empty lattice approximation (bottom frame) on the upper end of the 10-100~eV energy range. We observe significant differences in both the density of the bands and their gradients ($\partial E/\partial {\bf k}$: the group velocity). As these deviations would affect the scattering rates, it is likely that they would also affect the charge-carrier thermalization rate. For accurate treatment, then, we use the DFT bands.

\subsection{\label{sec:method.phonon}Phonon Dispersion}
We utilize density functional perturbation theory (DFPT) within QE to evaluate the phonon dispersion. It calculates the lattice dynamical matrix for a given perturbation, $\bf q$, which can then be diagonalized to obtain the eigenvalues, $\omega_{{\bf q}}^{\eta}$, for each of the possible vibrational modes for a given crystal. Eigenstates are obtained for a set of 50 points in the irreducible wedge corresponding to an $8\times8\times8$ BZ mesh. The main purpose in obtaining the phonon dispersion is to evaluate the carrier-phonon interaction, which is covered below (see section \ref{sec:method.epi}). 
\begin{figure}[!]
\includegraphics[height=3.5in, width=3.5in]{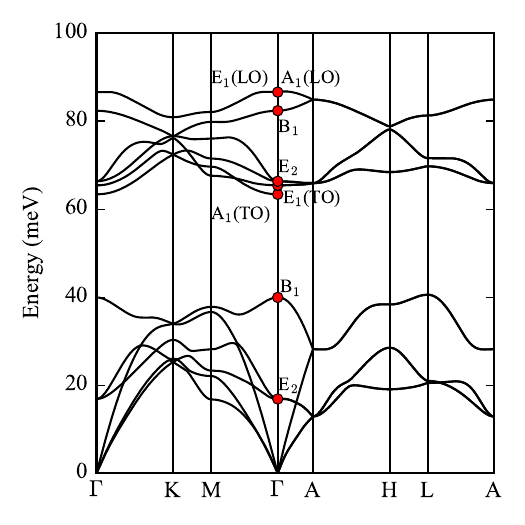}
\caption{\label{fig:ph_disp}Phonon dispersion of wurtzite GaN along several symmetry lines of the BZ. Several points at $\Gamma$ are indicated with red dots and labels. These energies are compared to experiment in Table~\ref{tab:table2}.}
\end{figure}

We plot the phonon dispersion calculated along several symmetry lines in the BZ in Fig.~\ref{fig:ph_disp}. The lowest three branches are acoustic, with the first two and the third known as transverse and longitudinal acoustic (TA and LA), respectively. The other nine branches are optical; all but the highest branch, which is longitudinal optical (LO), are transverse optical (TO). 

The highest energy of the LO branch at $\Gamma$ (E$_{1}$ (LO)) is 86.73~meV, which slightly underestimates the value measured by inelastic x-ray scattering of 90.4~meV~\cite{Ruf01} and that measured by Raman scattering of 92.1~meV ~\cite{Azuhata95}. In fact, in Table~\ref{tab:table2}, several points on the phonon spectrum can be compared to those measured in Refs.~\cite{Ruf01,Azuhata95}: Whereas the first six branches seem to be in good agreement, the energies of the highest six are consistently underestimated by $\sim$~3-5~meV. DFPT calculations from Refs.~\cite{bungaro_2000,karch_1998} and frozen phonon calculations from Ref.~\cite{gorczyca_1995} are also included in Table~\ref{tab:table2}. We observe deviations among these theoretical works up to $\sim$~5~meV, and deviations of similar magnitude are seen when comparing Refs.~\cite{bungaro_2000,karch_1998,gorczyca_1995} with the experimental data. We conclude, therefore, that the deviations observed between our work and experiment are in line with the expected accuracy of DFPT calculations and other theoretical results.

\subsection{Carrier-Phonon Interaction}
\label{sec:method.epi}
In QE, the carrier-phonon interaction can be investigated by calculating the electron-phonon matrix elements using an included package called EPW (Electron-Phonon coupling using Wannier functions)~\cite{Ponce16,Giustino07}. This program utilizes a Wannier-Fourier interpolation scheme to obtain the matrix elements on an arbitrarily fine mesh. The resulting matrix elements are given in the following form:
\begin{equation}
g_{nn'}^{\eta}({\bf k},{\bf k'})=\left\langle \psi_{n'}({\bf k'})\left|\frac{\partial V_{\rm eff}}{\partial \bf q}\right|\psi_{n}({\bf k}) \right\rangle, 
\label{eq:matrix_el}
\end{equation}
where $\psi_{n}({\bf k})$ is the initial electronic state, $\psi_{n'}({\bf k'})$ is the final state upon scattering with a phonon of wave vector $\bf q$ and mode $\eta$, and $V_{\rm eff}$ is the converged self-consistent potential from DFT.

We calculate the electron-phonon matrix elements for uniform grids of $\bf k$ and $\bf q$ for all valence bands (18 in total) and for the first 17 conduction bands. The reason for using only the first 17 conduction bands will become apparent in section \ref{sec:scattering.ELR.results}, where the rates for all scattering mechanisms are plotted together. In short, the rates of other scattering types (impact ionization and plasmon emission) exceed those of phonon scattering by a significant margin for intermediate energies (above $\sim$~10~eV). Therefore, it is unnecessary to perform this calculation for energies much higher than 10 eV. We note, also, that as GaN is a polar material, we include in the calculation the long-range contributions to the carrier-phonon interaction, from polar optical phonons.

\section{\label{sec:scattering}Scattering Mechanisms}
With the material properties determined, we now turn to the evaluation of the relevant charge-carrier interaction rates in GaN. For the study of hot-electron thermalization, we have chosen to consider phonon scattering, impact ionization, and valence-electron plasmon scattering. We will first discuss phonon scattering, and then we will introduce the energy-loss rate, which includes scattering from both impact ionization and plasmons.
\begin{figure}[!]
\includegraphics[height=5in, width=3.5in]{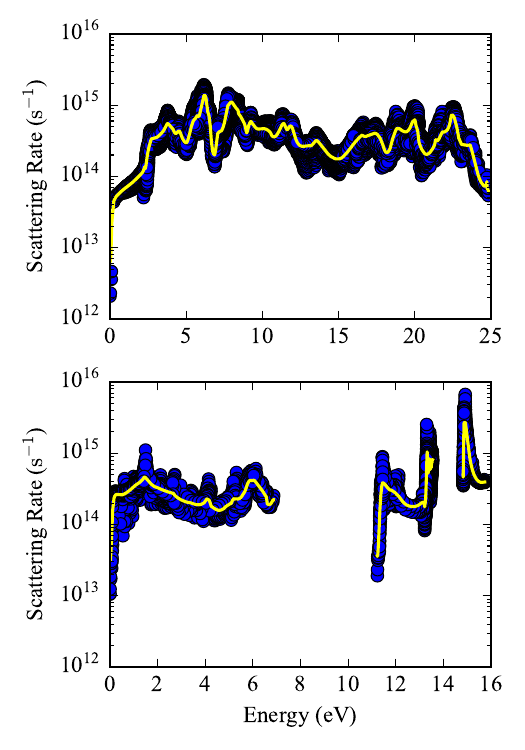}
\caption{\label{fig:ph_rates}Top: The total electron-phonon scattering rates in bulk wurtzite GaN. The rates are first plotted for many k points on the BZ mesh to observe the spread at a given energy and, thus, the k-dependence of the rates (blue markers). Over these points, the scattering rate averaged over equi-energy shells is plotted (yellow solid line). Bottom: The total hole-phonon scattering rate for holes in all 18 valence bands (including {\it d} bands).}
\end{figure}

\subsection{\label{sec:scattering.ph}Carrier-Phonon Scattering}
To calculate the scattering rate between a carrier of wave vector {\bf k} in band $n$ and a phonon of wave vector {\bf q} of branch $\eta$, we use first-order time-dependent perturbation theory (Fermi’s golden rule):
\begin{multline}
\frac{1}{\tau_{\eta}({\bf k},n)} \approx \frac{2\pi}{\hbar} \sum_{n',{\bf q}}\left| 
    g_{nn'}^{\eta}({\bf k},{\bf k}') \right|^{2} \left( N_{\bf q}+\frac{1}{2}\mp\frac{1}{2} \right ) \\
           \times \delta\left[E_n({\bf k})-E_{n'}({\bf k}')\pm\hbar\omega_{\bf q}^\eta\right],
\label{eq:FGR_EPW}
\end{multline}
with
\begin{equation}
N_{\bf q}=\frac{1}{e^{(\hbar\omega_{\bf q}/k_{\rm B}T)}-1}.
\label{eq:bose_einst}
\end{equation}
Here $g_{nn'}^{\eta}({\bf k},{\bf k}')$ is the electron-phonon matrix element provided by EPW (section \ref{sec:method.epi}), ${\bf k}'={\bf k}\pm{\bf q}$ is the ``final'' wave vector of the scattered carrier, which has been mapped back into the first BZ, and $n'$ is the final band of the scattered carrier. The numerical integration over the delta function is evaluated using a similar method to that reported by Fischetti and Laux \cite{MVF88}. This approach involves the use of the discretized BZ and identifying momentum- and energy-conserving cubes in the mesh. For such cubes, we employ Bl{\"o}chl's tetrahedron method to calculate the DOS, which is plugged into the summation of Eq.~(\ref{eq:FGR_EPW}) in place of the delta function.

The electron-phonon scattering rates, evaluated for the lowest 17 conduction bands, are shown in Fig.~\ref{fig:ph_rates} (top frame). The markers (blue) represent the rates calculated at 3234 points in the irreducible wedge. Over these markers, we have also plotted a solid line representing the average of the rates over equi-energy shells (yellow). At each energy, a spread, about the average, is observed in the rates, indicating that for much of this energy range, the rates have a relatively strong {\bf k}-dependence. This {\bf k}-dependence suggests that a purely energy-dependent rate is not sufficient to accurately simulate the scattering process in a FBMC simulation. Overall, the general shape and magnitude of the average compares well with those of Bertazzi {\it et al.} \cite{Bertazzi09}, who used an empirical pseudopotential method to generate the electronic band structure, the linear response method within density functional perturbation theory to obtain the phonon dispersion, and a deformation potential to produce the scattering rates.

The hole-phonon scattering rates are shown in the bottom frame of Fig.~\ref{fig:ph_rates}. Bertazzi {\it et al.} also calculated the hole-phonon rates, but only for the highest six valence bands, corresponding to energies up to $\sim$~7-8~eV. Over this energy range, the general shape and magnitude of our hole-phonon scattering rates are in agreement with Ref.~\cite{Bertazzi09}. 

\subsection{\label{sec:scattering.ELR}Impact Ionization and Plasmon Scattering}
While electron- and hole-phonon scattering dominate in the low-energy regime, at higher energies, impact ionization and plasmon scattering become possible and more prevalent. From an experimental perspective, information about these scattering mechanisms is typically obtained via electron energy loss spectroscopy (EELS). The majority of the peaks and shoulders that appear in the resulting spectrum represent single electron excitations (impact ionization), while a few represent collective excitations of the valence electrons (valence-electron plasmons). These collective excitations, or plasmons, decay via Landau damping \cite{Landau46}, which results in the creation of an EHP. We note that for both impact ionization and plasmon emission, we ignore excitons.

The total rate at which carriers scatter by these two mechanisms can be calculated using the dielectric function. This is so because the imaginary part of the inverse dielectric function (also known as the energy-loss function) is directly related to the EEL cross section. Here we include only plasmon emission, as the equilibrium plasmon occupation number is effectively zero ($N(\omega)=0$). In Chapter 2.6 of reference \cite{Nozieres99}, FGR and the dissipation-fluctuation theorem \cite{Nyquist28,Callen51} are used to express the equilibrium scattering rate via the dielectric function as
\begin{multline}
\frac{1}{\tau_{n}^{\rm (ELR)}({\bf k})} = 
   \frac{2\pi}{\hbar} \sum_{n'} \int\frac{{\rm d}{\bf q}}{(2\pi)^3} \frac{e^{2}\hbar}{q^2} 
       \int\frac{{\rm d}\omega}{2\pi} \ {\rm Im} \left[\frac{-1}{\varepsilon({\bf q},\omega)}\right] \\
           \times \delta \left [ E_n({\bf k})-E_{n'}({\bf k}+{\bf q}) \pm \hbar\omega \right]. 
\label{eq:ELR}
\end{multline}
Here ${\rm Im}\left[{-1}/{\varepsilon({\bf q},\omega)}\right]$ is the energy-loss function, and $\omega$ and $\bf q$ are the frequency and wave vector, respectively, of the resulting EHP or plasmon. We call this scattering rate the carrier energy-loss rate (ELR).
\begin{figure}[!]
\includegraphics[height=5.0in, width=3.5in]{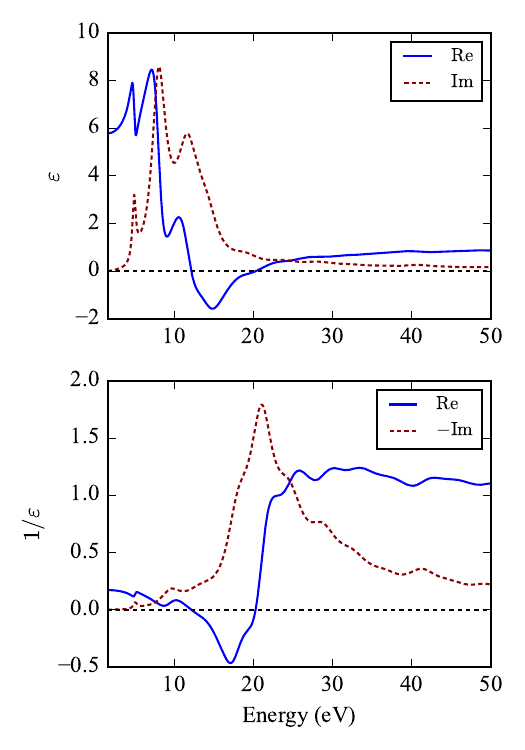}
\caption{\label{fig:eps_smallq}Top: The real and imaginary parts of the dielectric function for ${\it q}=0.0023 1/{\rm \AA}$. Bottom: The real and imaginary parts of the inverse of the dielectric function for ${\it q}=0.0023 1/{\rm \AA}$.}
\end{figure}
\begin{figure*}
\includegraphics[height=3in, width=7in,trim={0.9mm 0 0 0.9mm},clip]{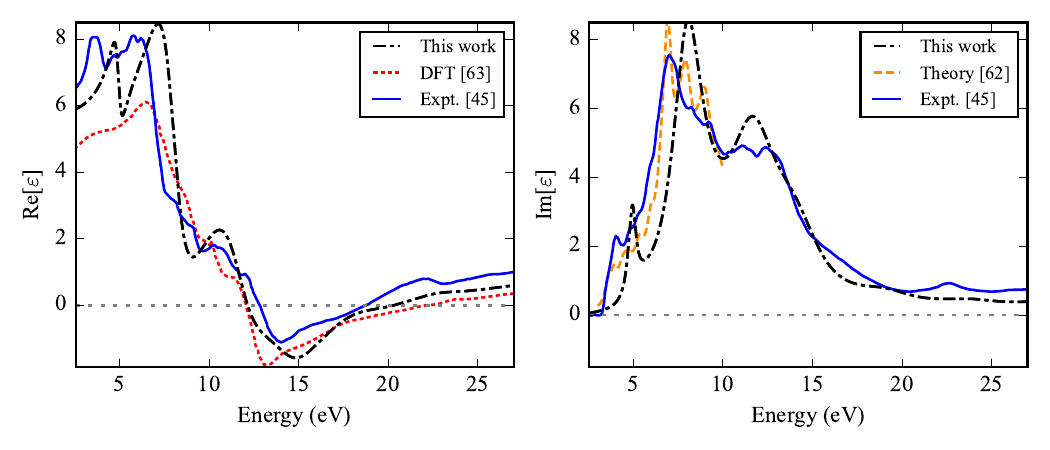}
\caption{\label{fig:eps_brockt}The real (left frame) and imaginary (right frame) parts of the dielectric function as calculated by time-dependent DFT (dash-dot) with the calculated results of Refs.~\cite{BENEDICT99,eljarrat_2016} and the experimental results of Refs.~\cite{Brockt00}.}
\end{figure*}

\subsubsection{\label{sec:scattering.ELR.diel}Calculation of the Dielectric Function}
A number of techniques exist to obtain information about the EEL spectrum, and thus the loss function. In this work, we use time-dependent DFT via the code known as {\it turboEELS} \cite{Timrov15}, which is included in the QE package. It utilizes linearized time-dependent DFT within the Liouville-Lanczos approach to optical spectroscopies to calculate the EEL spectrum. First, we perform a standard ground-state DFT calculation with a $12\times12\times12$ Monkhorst-Pack grid. The result is then passed into {\it turboEELS} along with the wave vector, {\bf q}, associated with the momentum transferred in the excitation process.  Additionally, we use 1000 Lanczos iterations with a bi-constant extrapolation of the Lanczos coefficients up to 20000 and a Lorentzian broadening of 0.02 eV.

We repeat this calculation for a set of  {\bf q} and $\omega$, so that an interpolation of the dynamic dielectric function ($\varepsilon({\bf q},\omega)$) can be performed for any $({\bf q},\omega)$. Ideally, one would obtain these results on a mesh of {\bf q} points, spanning the irreducible wedge of the BZ. This calculation, however, is a difficult task due to the computational cost of the calculation for each {\bf q}. Instead, the dielectric function is calculated for a set of points along the first crystal axis (perpendicular to the c-axis) of the BZ. All dielectric function results shown below are from this data set. In section \ref{sec:scattering.ELR.qdep}, for a comparison, the dielectric function is also calculated for several points along the third crystal axis (parallel to the c-axis).  Eq.~(\ref{eq:ELR}) is then solved separately for each data set (perpendicular and parallel to the c-axis), assuming isotropy of the dielectric function. While the dielectric function is not isotropic for GaN, these two calculations can be used to provide an idea of how strongly the ELR depends on the anisotropy of the loss function.

We first calculate $\varepsilon({\bf q},\omega)$ for a small momentum transfer, ${\bf q}=(0.001,0,0)$, which corresponds to a magnitude of 0.0023 $1/{\rm \AA}$ (Fig.~\ref{fig:eps_smallq}). The real and imaginary parts of $\varepsilon(\omega)$ (Fig.~\ref{fig:eps_smallq}, top frame) have been measured by spectroscopic ellipsometry~\cite{Goldhahn04,Kawashima97,Cobet09,BENEDICT99}, and via reflectivity~\cite{Lambrecht95} and EELS measurements~\cite{Brockt00} using the Kramers-Kronig transformation. Looking first at Im[$\varepsilon$], our results yield three major peaks at approximately 5, 8 and 11.5 eV. In Fig.~\ref{fig:eps_brockt} (right frame) we show a direct comparison with experimental and theoretical works. For the EELS data of  Ref.~\cite{Brockt00}, major peaks occur at 4, 7 and between 10.5-13 eV, roughly corresponding to those from this work. Additional smaller peaks occur at $\sim$~8, 9, 11 and 12.5 eV (in agreement with Refs.~\cite{Lambrecht95,Goldhahn04,Kawashima97,Cobet09,BENEDICT99}), which do not appear in our results. We see that first-principles calculations of Im$[\varepsilon]$ by Benedict {\it et al.}~\cite{BENEDICT99} were able to resolve the peaks at $\sim$~8 and 9~eV, but they do not include the sharp peak at $\sim$~4-5~eV.

For Re[$\varepsilon$] (left frame), we observe similar discrepancies between our work and experiment: Whereas our calculations seem to yield the major peaks, many of the more minor features do not appear in our results. We include a DFT calculation of Re[$\varepsilon$] from Ref.~\cite{eljarrat_2016} in Fig.~\ref{fig:eps_brockt} (left frame). We observe that at energies $<8$~eV the major peak magnitudes of Ref.~\cite{eljarrat_2016}, are significantly smaller than those of our work and Ref.~\cite{Brockt00}. Still, the peak locations seem to be in relatively good agreement. We note one key differences among these plotted data: the position of the zero of Re[$\varepsilon$] where the slope is positive. This zero occurs at $\sim$~20.3~eV for our work, and at $\sim$~22.9~eV for Ref.~\cite{eljarrat_2016}. Both are higher than experiment ($\sim$~18-19~eV), but our result is roughly 12\% closer. The importance of this zero will be discussed in the following section. Overall, we conclude that the accuracy of our calculations seems to be in line with the accuracy of other theoretical works.

The discrepancies between our calculations and measured dielectric function data most likely originate from the use of Lorentzian broadening in the {\it turboEELS} code. Broadening schemes, such as Gaussian and Lorentzian broadening, tend to smooth the results of a BZ integration, causing the details of the desired spectrum/data to be lost. In principle, one may adjust the broadening width to attempt to resolve these missing peaks, which we have done without success.  Another option (that avoids changing the BZ integration scheme) is to increase the density of the Monkhorst-Pack grid. For a few $\bf q$ points, we have incrementally increased the density beginning from $6\times6\times6$ and found convergence by $12\times12\times12$. For further improvement one would need to change the integration scheme to a more accurate numerical technique such as the tetrahedron method~\cite{Blochl94} or Gilat-Raubenheimer method~\cite{gilat_1966,Liu17}. Such an endeavor, however, would require significant effort, as one would need to change the {\it turboEELS} source code. For simplicity, we use the results from {\it turboEELS} without making any changes.

\subsubsection{\label{sec:scattering.ELR.lifetime}Plasmon Lifetime and Dispersion}
Before calculating the carrier ELR, we first turn our attention to identifying the bulk plasmon peaks of wurtzite GaN in the loss function. In principle, by identifying these peaks, one may calculate the plasmon emission rate separately from the impact ionization rate. This approach allows one to treat the two phenomena separately within the FBMC code. The primary concern here is the lifetimes of the plasmons. As mentioned previously, plasmons decay via Landau damping, which results in the creation of an EHP. When the decay rate is sufficiently low, plasmons may live long enough in the system to be reabsorbed or slow the thermalization process. Dealing with these plasmons and their transport can significantly increase the complexity of the FBMC code.
\begin{figure}[!]
\includegraphics[height=5in, width=3.5in]{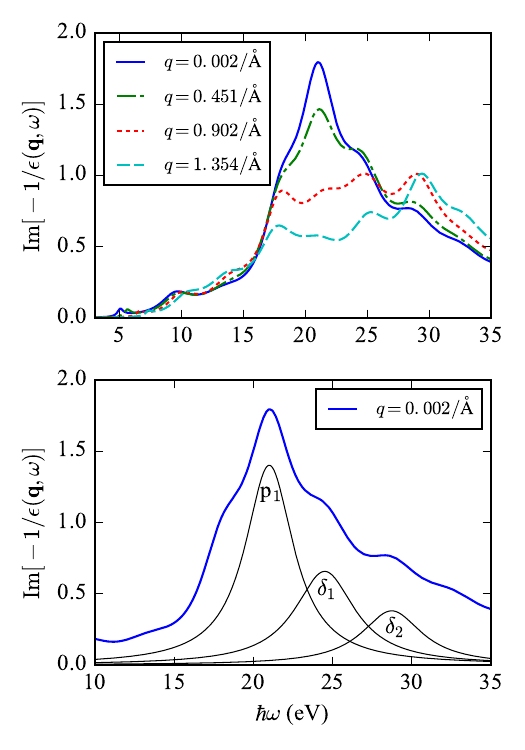}
\caption{\label{fig:lor_fit}Top: The loss function of GaN calculated using time-dependent DFT. The function Im$\left[ -1/\varepsilon({\bf q},\omega) \right]$ is plotted vs. the energy, $\hbar\omega$, for the indicated values of {\it q}. Bottom: The loss function for the indicated value of {\it q} with Lorentzian fits for the major {\it sp}-electron plasmon peak (p$_1$) and the two {\it d}-electron peaks ($\delta_1$ and $\delta_2$).}
\end{figure}

To identify plasmon peaks in the loss function, one may first employ the so-called weak damping approximation \cite{Hamann20}. Within this approximation, bulk plasmon energies can be found at zeroes of Re[$\varepsilon$] (where the slope is positive). Plasmons actually occur at complex zeros of the full complex dielectric function, but, when damping is weak, real zeros of the real part are a good approximation. When $q$ is small, this approximation tends to hold, but for larger $q$, it breaks down. Additionally, there may be plasmon peaks for which the damping is strong enough so that the real part never vanishes on the real axis (even for small $q$).
\begin{figure}[!]
\includegraphics[height=5in, width=3.25in]{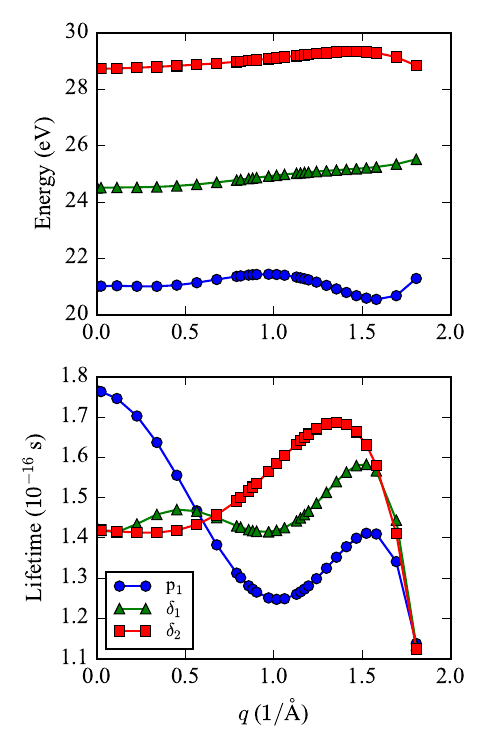}
\caption{\label{fig:disp_life}Plasmon dispersion (top) and lifetime (bottom) obtained from the position and width of the quasi-Lorentzian peaks (p$_1$, $\delta_1$, and $\delta_2$) shown in Fig.~\ref{fig:lor_fit}.}
\end{figure}
\begin{figure*}
\includegraphics[height=4.5in, width=6in]{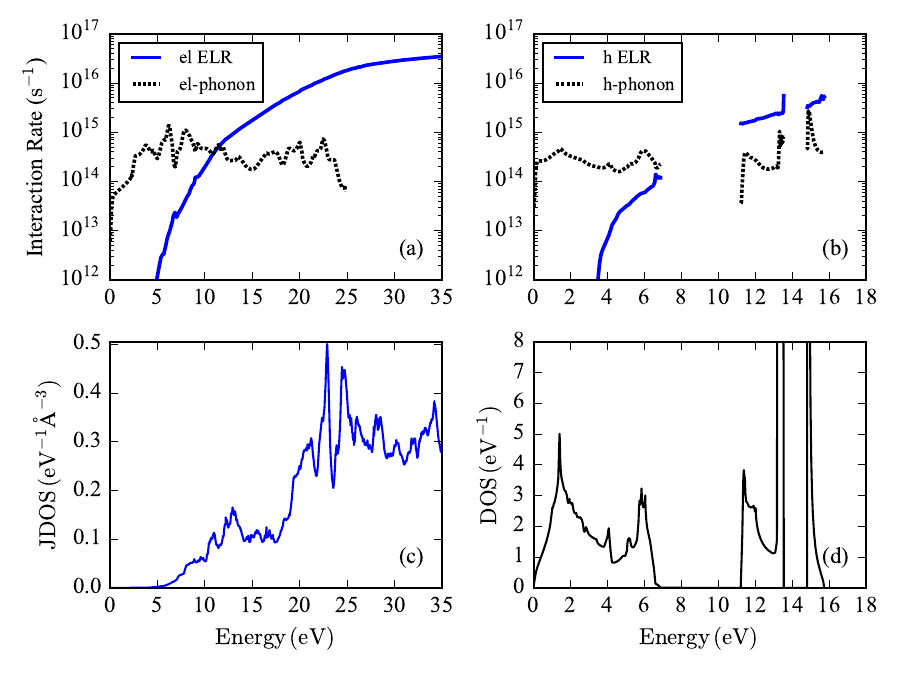}
\caption{\label{fig:ELR_valence band_DOS_JDOS}(a) The electron ELR (solid line) plotted with the electron-phonon scattering rate (dotted line). The rate flattens out at higher energies. (b) The hole ELR (solid) plotted with the hole-phonon scattering rate (dotted). Gaps correspond to gaps in the valence band (Fig.~\ref{fig:band_structure}). (c) The JDOS of GaN calculated using tetrahedron method. A significant increase is observed at $\sim$~17-18~eV, due to the onset of d-electron excitation. (d) Total DOS of the valence bands of GaN, calculated using tetrahedron method in QE. The large peaks of the {\it d}-band DOS are observed at $\sim$~13.5~eV and $\sim$~15~eV.}
\end{figure*}

In Fig.~\ref{fig:eps_smallq}, Re$[\varepsilon]$ goes through zero at $\sim$~20.3~eV, which corresponds almost exactly with the largest peak in the loss function. We call this the primary plasmon peak (p$_1$). It represents the major bulk plasmon predicted by the electron gas model,
\begin{equation}
\omega_{\rm p}^2 = \frac{e^2 n}{\varepsilon_0 m} \ ,
\end{equation}
to be at an energy of $\sim$~21.7~eV. Here $n$ is the valence electron density ($s$ and $p$ only) and $m$ is the free-electron mass. It originates from oscillations of the {\it s} and {\it p} valence electrons \cite{Dhall17}. EELS data from Refs.~\cite{Brockt00,Dhall17,Benaissa21,Sanchez04} yield a zero at a lower energy of $\sim$~18-19~eV (Fig.~ \ref{fig:eps_brockt}, left frame). In addition to p$_1$, two other peaks ($\delta_1$ and $\delta_2$) are observed at $\sim$~24 and $\sim$~28~eV (Fig.~\ref{fig:lor_fit}). The locations of these peaks are in closer agreement with experiment \cite{Brockt00,Dhall17,Benaissa21,Sanchez04}. In a study of the loss functions of AlN and GaN, Dhall {\it et al.}~\cite{ Dhall17} found that $\delta_1$ and $\delta_2$ are associated with collective excitations (plasmons) of the {\it d} electrons of Ga, and not single-particle excitations (impact ionization). As there are no zeroes of Re$[\varepsilon]$ on the real axis at $\sim$~24 and $\sim$~28 eV (Fig.~\ref{fig:eps_smallq}), one would not be able to identify these plasmons using the weak damping approximation. 

With the plasmon peaks identified, we fit Lorentzian curves to p$_1$, $\delta_1$, and $\delta_2$, as shown in Fig.~\ref{fig:lor_fit} (bottom frame). Several more Lorentzian curves, not shown in the figure, are fit to the surrounding peaks to improve the overall fit. Using this curve fitting, we track as a function of $q$ the peak positions, which yield the plasmon dispersion, and the full width at half maximum, which yields the plasmon lifetimes (Fig.~\ref{fig:disp_life}). The top frame of Fig.~\ref{fig:disp_life} indicates that the plasmon energy of p$_1$ is actually $\sim$~21~eV and not  $\sim$~20.3~eV, where Re$[\varepsilon]$ vanishes. This divergence suggests that the damping of p$_1$ is not negligible. 

The bottom frame of Fig.~\ref{fig:disp_life} shows that the plasmon lifetimes are all of order 10$^{-16}$~s. We note that the lifetime for p$_1$ is almost 50\% larger than that of the delta peaks for small $q$. As $q$ increases, the lifetime of p$_1$ clearly decreases, suggesting that the damping increases. In contrast, the lifetime of the delta peaks is always short at small $q$. We conclude that these magnitudes are short enough to assume immediate plasmon decay in the FBMC simulation. Therefore, we avoid the complexity of having to treat the plasmons separately from impact ionization, as discussed above.

In the top frame of Fig.~\ref{fig:lor_fit}, the loss function has a strong dependence on {\it q}, especially for the energy range from $\sim$~16~eV to $\sim$~35~eV. The shape and magnitude of these peaks change considerably, suggesting that the $q$-dependence must be accounted for in the calculation of the ELR. 

\subsubsection{\label{sec:scattering.ELR.results}Electron and Hole Energy-Loss Rates}
With the dielectric function calculated, we now evaluate the carrier ELRs. The numerical integration over the delta function in Eq.~\ref{eq:ELR} is performed in a similar way as in section \ref{sec:scattering.ph}. It begins with a search of the BZ for energy-and momentum-conserving cubes \cite{MVF88}. Here, however, there is an additional integration over $\omega$. For each cube, we check for energy-conservation using a list of possible carrier energy losses, $\hbar\omega$ (from a discretization of the energy range from 0 to the kinetic energy of the incident carrier). If a given $\hbar\omega$ satisfies energy conservation in the cube, the DOS is calculated and included in the summation. 

In Fig.~\ref{fig:ELR_valence band_DOS_JDOS} (a) and (b), we plot the calculated total electron and hole ELR, respectively. The phonon scattering rates for each are also shown. As previously mentioned, for electrons, phonon scattering dominates up to $\sim$~10~eV above which the ELR dominates by 1-2 orders of magnitude. This fact makes the phonon interaction practically irrelevant above this point. For this reason, as discussed in section \ref{sec:method.epi}, electron-phonon matrix elements above the 17$^{\rm th}$ conduction band (25 eV) are not calculated. For the holes, phonon scattering is the major interaction mechanism up to the first gap at $\sim$~7~eV. Just above 6~eV, impact ionization (ELR) reaches a similar magnitude, and it dominates for all higher energies.

The sharp peaks in the hole-phonon scattering rate, observed at $\sim$~13.5~eV and $\sim$~15~eV, originate from the incredibly dense {\it d} bands seen in Fig.~\ref{fig:band_structure}. The DOS of these bands far exceeds the DOS of all other valence bands (Fig.~\ref{fig:ELR_valence band_DOS_JDOS} (d)) and even that of all conduction bands up to 100 eV. These {\it d}-band peaks lead to a significant jump in the JDOS at $\sim$~17-18~eV (Fig.~\ref{fig:ELR_valence band_DOS_JDOS} (c)), caused by the onset of {\it d}-electron excitation (approximately the primary energy gap (3.4 eV) plus the depth of the first {\it d} bands). This jump accounts for the rather large magnitude of the electron ELR (10$^{16}$-10$^{17}$ s$^{-1}$) in Fig.~\ref{fig:ELR_valence band_DOS_JDOS} (a), which will be discussed further in the following section. 

The large ELRs shown in Fig.~\ref{fig:ELR_valence band_DOS_JDOS} may arguably raise some questions about the validity of the first Born approximation, which assumes that the wavefunction of the incident carrier is not appreciably affected by the scattering potential~\cite{griffiths_1995}. This concern has been addressed by Quinn and Ferrel \cite{Quinn58,Quinn62}, Pines \cite{Pines_1956}, Penn \cite{Penn87}, and others who have calculated ELRs of the same magnitude. They have concluded that the electron energies at which the ELRs are high, are so large as to still render the broadening of electronic states acceptably small and that the use of perturbation theory is justified.

\subsubsection{\label{sec:scattering.ELR.altmethods}Alternative Methods for Calculating the Impact Ionization Rate}
\begin{figure*}
\includegraphics[height=2.5in, width=6in,trim={1.7mm 0 0 3mm},clip]{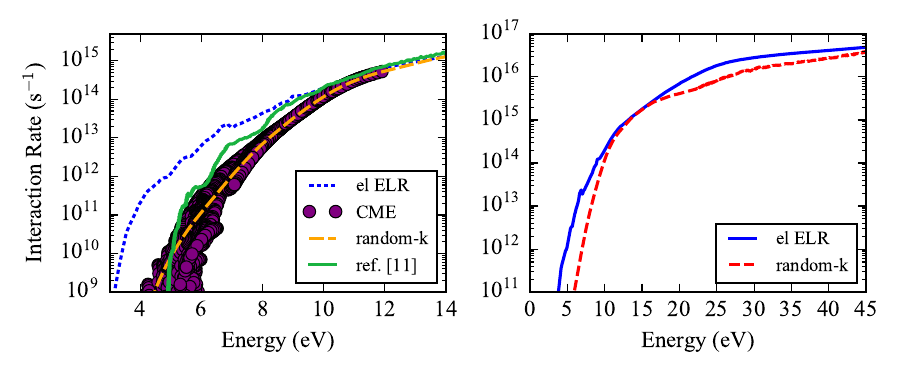}
\caption{\label{fig:II_all}Left: Impact ionization rates (calculated by both random-k approximation and the CME approximation), the electron ELR (calculated via the dielectric function), and the impact ionization rate found by Bertazzi {\it et al.} \cite{Bertazzi09}. These are plotted in the relatively low-energy regime. Right: Electron ELR and impact ionization rate via random-k up to higher energies.}
\end{figure*}
In order to identify separately collective (plasmon) and single-particle (impact ionization) contributions to the ELR obtained above, in this section, we present the carrier impact ionization rate (without plasmon scattering) calculated using two alternative methods to Eq.~(\ref{eq:ELR}). These include the so called ``constant matrix element'' (CME) approximation and Kane's random-k approximation \cite{Kane67}, which are derived from the first Born approximation \cite{Sano90,MVF96}. Here we plot the results together with the ELR, which helps to assess the accuracy of the calculated ELR and to identify the energy regions in which impact ionization dominates.

As one might expect, the major assumption of the CME approximation is that the matrix element associated with the scattering process, $M$, is a constant. We define this constant as the average value of the matrix element over all points in the BZ, all bands, and over both normal and {\it Umklapp} processes. This assumption emphasizes the idea that the matrix element plays a small role in determining the energy dependence of the ionization rate. Within this approximation, we write the impact ionization rate as \cite{MVF96}:
\begin{widetext}
\begin{equation}
\frac{1}{\tau^{\rm (CME)}_{\rm ii}({\bf k},n)} =
    \frac{2\pi}{\hbar} \langle M^2 \rangle \sum_{\bf G} \sum_{n_{\rm v}n_{\rm c}'n_{\rm c}''}
         \int\frac{{\rm d}{\bf k'}}{(2\pi)^3} \int\frac{{\rm d}{\bf p'}}{(2\pi)^3} \
                \delta \left[ E_n({\bf k})+E_{n_{\rm v}}({\bf p})-E_{n_{\rm c}'}({\bf k'})-E_{n_{\rm c}''}({\bf p'}) \right ] \ ,
\label{eq:II_CME}
\end{equation}
\end{widetext}
where
\begin{equation}
\langle M^2 \rangle = \frac{e^4a_0^4m^2}{(2\pi)^4\varepsilon_{\rm s}^2}  \ . \nonumber
\end{equation}

Here $\bf k$ is the wave vector of the incident carrier in band $n$, with energy $E_n({\bf k})$. The wave vector $\bf k'$ is the crystal momentum of the final state of the incident carrier, and $\bf p$ and $\bf p'$ are, respectively, the initial and final crystal momenta of the excited electron. As this excited electron begins in the valence band, we label the initial band as $n_{\rm v}$ (v for valence band). The bands $n_{\rm c}'$ and $n_{\rm c}''$ are the final bands of the incident and excited particles, respectively (c for conduction band). The matrix element, $M$, is the (anti)symmetrized screened Coulomb matrix element. For $M$, $a_0$ is the lattice constant, $\varepsilon_{\rm s}$ is the static dielectric constant, and m is a number of order 1.

For the random-k approximation, we take Eq.~(\ref{eq:II_CME}) and employ further simplifications. The major assumption (along with those of CME) is that the JDOS available to the scattered (incident and excited) particles primarily controls the kinematics of the ionization process rather than momentum conservation \cite{MVF96}. In other words, this assumption suppresses momentum conservation in the ionization process. It holds when {\it Umklapp} processes dominate the pair production channel, causing momentum randomization. Within random-k, Eq.~(\ref{eq:II_CME}) becomes:
\begin{widetext}
\begin{equation}
\frac{1}{\tau_{\rm rk}^{\rm (ii)}(E)} = \frac{2\pi}{\hbar} \frac{\Omega_{\rm c}}{8} \langle M^2 \rangle
     \int_{0}^{E-E_{\rm g}} {\rm d}E_{\rm c} \int_{0}^{E-E_{\rm g}-E_{\rm c}} {\rm d}{E_{\rm v}} \ 
           \mathcal{D}_{\rm c}(E_{\rm c}) \ \mathcal{D}_{\rm v}(E_{\rm v}) \ \mathcal{D}_{\rm c}(E-E_{\rm g}-E_{\rm c}-E_{\rm v}).
\label{eq:II_rand_k}
\end{equation}
\end{widetext}
Here $\Omega_{\rm c}$ is the volume of the crystal primitive cell, $E_{\rm c}$ is the energy of the “partner” electron (measured from the conduction band minimum) after jumping across the gap, $E_{\rm v}$ is the energy of the “secondary” hole, $E$ is the initial energy of the “primary” particle, and $\mathcal{D}_i$ ($i$='c' or 'v') is the total DOS at a given energy in either the valence band or conduction band.
\begin{figure*}
\includegraphics[height=2.5in, width=6in,trim={2mm 0 0 4mm},clip]{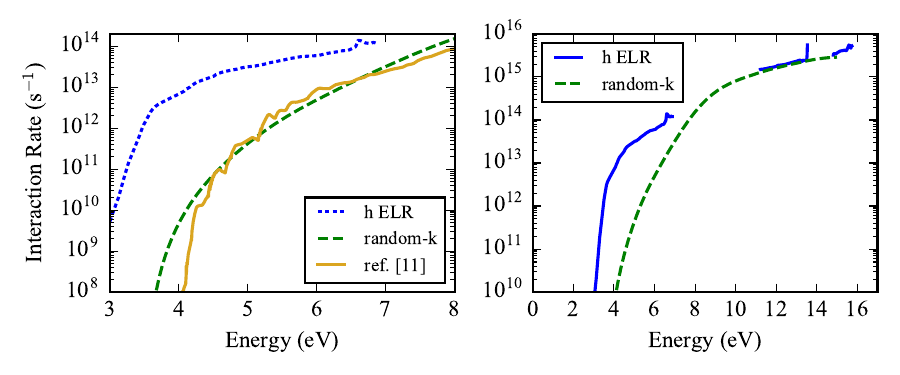}
\caption{\label{fig:II_all_h}Left: The hole impact ionization rate (calculated by both random-k approximation), the hole ELR (calculated via the dielectric function), and the impact ionization rate found by Bertazzi {\it et al.} \cite{Bertazzi09}. This plot covers the lower-energy range. Right: The Hole ELR and impact ionization rate via random-k up to higher energies. Convergence is observed at higher energies.}
\end{figure*}

The CME and random-k results for electrons are shown in Fig.~\ref{fig:II_all} (left frame). The CME rates are plotted for several {\bf k} points spanning the irreducible wedge. As with phonon scattering (Fig.~\ref{fig:ph_rates}), we observe a spread in the rates, corresponding to momentum dependence. As energy increases, the spread narrows and converges to the random-k line. This convergence suggests that  momentum dependence weakens for increasing energy and {\it Umklapp} processes become the dominant scattering process. A similar spread is observed in the work of Bertazzi {\it et al.} \cite{Bertazzi09}, who used an MC approach to solve the scattering rate equation within the Born approximation, considering both energy and momentum conservation, and including a dynamic matrix element. In Fig.~\ref{fig:II_all}, we show only the average ionization rate from Ref.~\cite{Bertazzi09}, but the convergence to the random-k line is evident.

Additionally, we include in both frames of Fig.~\ref{fig:II_all} the electron ELR. It too converges to the random-k results as energy increases. Between 15 and 20 eV, the ELR diverges from the random-k line as, it seems, plasmon emission begins to dominate. Eventually the two converge again at $\sim$~45~eV, suggesting that impact ionization is again the major scattering mechanism. For low energies (below $\sim$~10~eV), we note that the electron ELR data are orders of magnitude larger than the others. In 1956, Pines proposed the existence of acoustic plasmons that may also appear as excitations in the loss function \cite{Pines1956a}. It is possible that interactions with acoustic plasmons may be responsible for the discrepancy in question. From a practical perspective, this discrepancy is irrelevant as phonon scattering dominates in this energy regime. Indeed, we have artificially modified the ELR to more closely match the rates of Bertazzi {\it et al.} and found that this does not affect the MC simulation results in any significant way.

We note also that this low-energy discrepancy may be at least partly due to a numerical artifact in the Im$[-1/\varepsilon({\bf q},\omega)]$, resulting from the use of Lorentzian broadening in {\it turboEELS}. Near the gap energy (3.4 eV), the loss function should go to zero, but instead, it remains nonzero down to much lower energies. As a result, Im$[-1/\varepsilon({\bf q},\omega)]$ is likely too large in this energy region causing a spurious increase in the ELR.

Interestingly, the random-k rates reach the same order of magnitude as the ELR ($10^{16}$-$10^{17}$ s$^{-1}$; $E>\sim30$~eV). Because the random-k approximation assumes, by definition, that the impact ionization scattering process is controlled primarily by the JDOS, it can be concluded that the JDOS is the primary factor in driving the random-k results to this magnitude. This conclusion supports the claim, in the previous section, that the ELR are driven to such high magnitudes by the jump in the JDOS caused by {\it d}-electron excitations and the extremely large DOS of the {\it d} bands.

For holes (Fig.~\ref{fig:II_all_h}, left frame), we observe similar phenomena. All plotted rates converge to the random-k results as the energy increases. Additionally, we observe the same discrepancy between the hole ELR and the hole random-k results at low energies. We, again, attribute this to the possibility of acoustic plasmon scattering with a contribution from numerical artifacts in the calculation of the loss function. In practice, just as with electrons, phonon scattering dominates in this relatively low-energy regime, so there is no need to make any changes or improvements at this time.

\subsubsection{\label{sec:scattering.ELR.qdep}Dependence of the ELR on the Direction of \textbf{q}}
\begin{figure}[!]
\includegraphics[height=3in, width=3.5in]{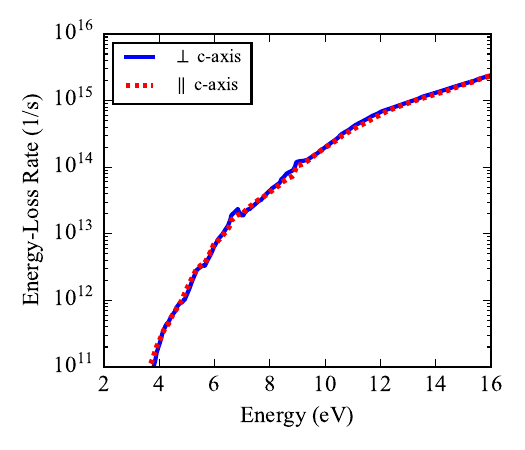}
\caption{\label{fig:b1_b3_comp}The electron ELR calculated using the loss function obtained along a crystal axis perpendicular to the c-axis (solid line; blue) and that obtained parallel to the c-axis (dashed line; red). Good agreement is observed, suggesting that the anisotropy of the dielectric function does not significantly affect the ELR, allowing one to ignore it.}
\end{figure}

Lastly, in this section, we come back to the issue of the anisotropy of the dielectric function in GaN and its effect on the ELR. As mentioned above, we have calculated the ELR using the dielectric function calculated perpendicular to the c-axis (``ELR $\perp$'', see Figs.~\ref{fig:II_all} and \ref{fig:II_all_h}) and that calculated parallel to the c-axis (``ELR $\parallel$''). In Fig.~\ref{fig:b1_b3_comp}, the two are plotted together for comparison. It is clear that there is excellent agreement between these rates. We conclude, then, that for the purposes of this work the anisotropy of the dielectric function does not significantly affect the ELR, and isotropy may be assumed without losing accuracy.

\section{\label{sec:FBMC}Full-Band Monte Carlo Simulation}
In this section, we present our FBMC simulation. For the general setup, the use of a synchronous ensemble, and the scattering mechanism selection, we have followed the methods laid out by Jacoboni and Reggiani \cite{Jacoboni83}. For the full band inclusion, we have followed Fischetti and Laux \cite{MVF88}. To establish the accuracy of the FBMC code and to further test the DFT data, we first calculate the low- and high-field transport characteristics. We then move to the simulation of the full thermalization of 100-eV electrons and the generated EHPs.

\subsection{\label{sec:FBMC.init_FF}Particle Initialization and Carrier Free Flight}
To begin the MC simulation, we define the initial states of a set of charge carriers (electrons and/or holes). For the low- and high-field transport, these carriers are first assigned a random energy under the Fermi-Dirac distribution, using the rejection technique~\cite{footnote_rejtech,Jacoboni83}. For the hot-carrier thermalization, energies are assigned under a Gaussian distribution centered at 100 eV. Wave vectors are chosen, for each, by scanning the BZ mesh for all cubes that intersect the constant-energy surface $E_i$, of the i$^{\rm{th}}$ charge carrier. Each of these cubes is given a weight equal to the DOS at $E_i$ (via the tetrahedron method). A cube is then selected randomly by the rejection technique, using the weight as a probability distribution. A {\bf k} within the cube is then chosen, such that $E({\bf k})\approx E_i$. This selection is done by creating a ``sub-mesh'' within each cube and interpolating the energies at each sub-mesh point, beforehand, and storing them in a look-up table. The point with the closest energy to $E_i$ is selected.

After initialization, the carriers enter a ``free flight''. For this calculation, we employ a synchronous ensemble technique~\cite{sync_ens,Jacoboni83,MVF88}. We handle the free flight and the process of updating the particle state as is done in Ref.~\cite{MVF88}. The one major exception is that we use a variable time step, d$t$. The magnitude of d$t$ changes with each time step, as the maximum scattering rate changes throughout the simulation. We update d$t$ at the beginning of each free flight by obtaining the total scattering rate for each carrier and finding the maximum. From there, d$t$ is assigned a value equal to the inverse of the maximum scattering rate divided by 10. This approach is chosen due to the orders-of-magnitude change of the scattering rate over the course of the thermalization process. If one were to fix the time step based solely on the scattering rate at the beginning, it would take an exorbitantly long time to run the full thermalization process. This method is significantly faster and more efficient, but it would lose its efficiency if the scattering-rate magnitude changes very little during a simulation.

\subsection{\label{sec:FBMC.scattering}Particle Scattering and the Selection of Final States}
Following free flight, we determine whether each carrier scatters, using the techniques described in Ref.~\cite{Jacoboni83}. Once a scattering mechanism is chosen (or not), a final state of the incident carrier and, if applicable, the states of the excited carriers are selected.

Selecting a final state after a carrier-phonon scattering event proceeds much in same way that the scattering rate is calculated in section \ref{sec:scattering.ph}. Following Ref.~\cite{MVF88}, we find all momentum- and energy-conserving cubes in the BZ (labeled with an index $j$), for all final bands $n'$. For each of these cubes, we calculate the density of final states $DOS_{jn}$ at the final energy and  
assign a weight: $P({\bf q}_j,n')=\left|g_{\eta} \left({\bf k},{\bf q}_j,n,n' \right)\right|^2 (N_{{\bf q}_j}) DOS_{j,n'}(E_n({\bf k})\pm\hbar\omega_{{\bf q}_j}^\eta)$, which is saved to a list. Using $P({\bf q}_j,n')$, one cube is chosen by rejection technique. A wave vector ${\bf k}'$ within the cube is then chosen, using the sub-mesh described in the previous section (\ref{sec:FBMC.init_FF}), such that $E_{n'}({\bf k}')\approx E_n({\bf k})\pm\hbar\omega_{{\bf q}_j}^{(\eta)}$.

For plasmon emission, because we have assumed immediate decay of emitted plasmons into EHPs, the final state selection process is identical to that of impact ionization. This process is treated in two steps: first, the emission of a plasmon/generation of a pair, and second, the division of the EHP momentum and energy between the excited electron and hole.

For the first part, we employ a variation of the technique used for phonon scattering (Ref.~\cite{MVF88}). We begin by taking the initial wave vector of the incident carrier, {\bf k}, and calculating ${\bf q}_j={\bf k}-{\bf k}_j$, where $j$ is the index for each cube in the BZ mesh, and {\bf k}$_j$ is the point at the center of cube $j$.  We then generate a list of possible energy losses, $\hbar\omega$. This list should span the energy range on which the dielectric function is calculated. In this case, the dielectric function is calculated from 0 to 100 eV, and a list of 200 energies is generated on this interval. For each of these energies, $E_n({\bf k})-\hbar\omega$ is calculated. The BZ is then searched for each cube in each band ($n'$) containing $E_n({\bf k})-\hbar\omega$. For these cubes, the density of states $DOS_{j,n}$, at the final energy and a probability 
$P({\bf q}_j,n',\omega)=(1⁄\left|{\bf q}_j\right|^2 ){\rm Im}\left[-1/\varepsilon({\bf q}_j,\omega )\right ] DOS_{j,n'}(E_n({\bf k})-\hbar\omega)$ 
are calculated and saved to a list. The rejection technique is then utilized, as before, to select a cube and an energy loss, $\hbar\omega$. The final wave vector of the incident particle is chosen, using the sub-mesh.

The final output of this section of code is the final state of the incident particle (energy and wave vector) and the energy and momentum of the EHP. 
To divide the pair energy and momentum between the excited electron and resulting hole, we utilize the random-k approximation. We consider a number of possible combinations ($E_{\rm el},E_{\rm h}=(\hbar\omega-E_{\rm g}-E_{\rm el})$) over the range from one extreme (electron absorbs all of the pair energy: $E_{\rm el}=(\hbar\omega-E_{\rm g})$) to the other ($E_{\rm el}=0$). Here $E_{\rm el}$ and $E_{\rm h}$ are measured from the conduction band minimum and valence band maximum, respectively. For each combination, the probability $P(E_{\rm el},E_{\rm h} )=\mathcal{D}_{\rm c} (E_{\rm el})\mathcal{D}_{\rm v} (\hbar\omega-E_{\rm el}-E_{\rm g} )$ is calculated and saved in a table. Here $\mathcal{D}_{\rm c} (E)$ and $\mathcal{D}_{\rm v}(E)$ are the total density of states at $E$ of the conduction bands and valence bands, respectively. The rejection technique is then used to select a combination ($E_{\rm el},E_{\rm h}$). With the energies selected, one simply finds all cubes that contain each energy and selects one, using rejection technique with the DOS as the probability distribution. Wave vectors within each cube are chosen using the sub-mesh method.

\subsection{\label{sec:FBMC.trans_char}Low- and High-Field Transport Characteristics}
\begin{figure}[!]
\includegraphics[height=3.5in, width=3.5in,trim={2.5mm 0 0 3mm},clip]{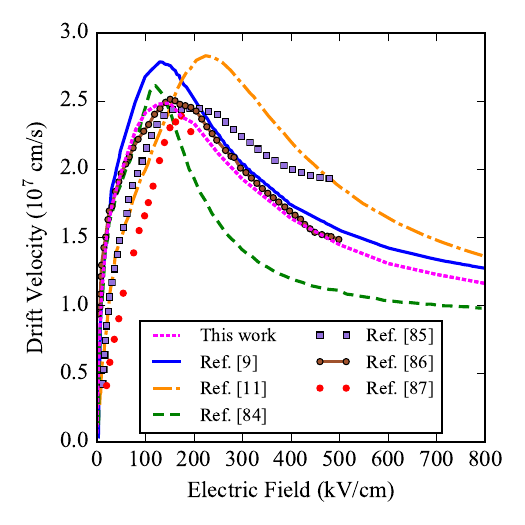}
\caption{\label{fig:dv_vs_F_curve}Average electron drift velocity as a function of applied electric field along the $\Gamma$-K symmetry line. This work (finely dashed line: magenta) is compared to the work in references \cite{Fang19,Bertazzi09,Wu05,Albrecht98,Yamakawa09,Barker02}.}
\label{fig:el_dv}
\end{figure}

\begin{figure*}
\includegraphics[height=2.75in, width=6in]{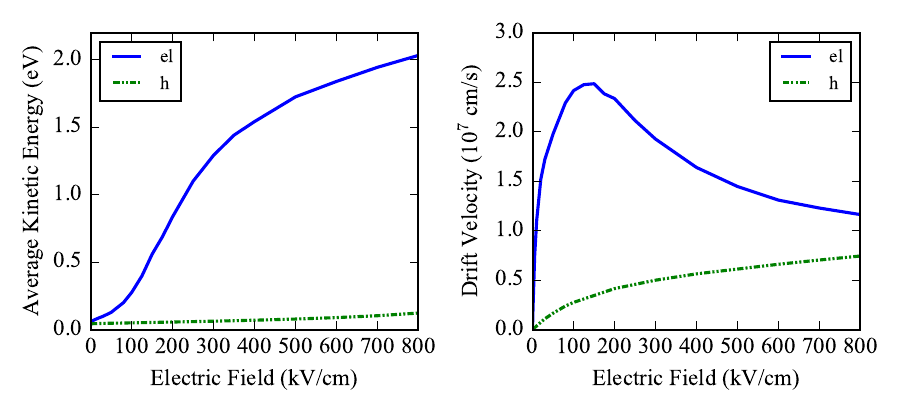}
\caption{\label{fig:el_h_E_dv}Left: Average electron and hole kinetic energy as a function of electric field from the MC simulations. Right:  Average electron and hole drift velocity as a function of electric field along the $\Gamma$-K symmetry line.}
\end{figure*}

As mentioned above, to establish the accuracy of the FBMC code and to further test our DFT results, we first calculate the low- and high-field transport characteristics. We present these data in Figs.~\ref{fig:el_dv} and \ref{fig:el_h_E_dv}. Fields are directed along the $\Gamma$-K symmetry line, for transport on the basal plane. In Fig.~\ref{fig:el_dv}, we plot the calculated average electron drift velocity of this work with the results of a few theoretical works \cite{Fang19,Bertazzi09,Wu05,Albrecht98,Yamakawa09} and one experimental \cite{Barker02}. In all cases, we see a similar trend: a relatively steep slope at low fields followed by a peak, a gradual decrease, and eventual saturation. We extract a peak velocity of approximately $2.5\times10^7$ cm/s, which agrees with Refs.~\cite{Yamakawa09,Albrecht98} and is in relatively good agreement with the experimental result: $\sim 2.4\times10^7$ cm/s \cite{Barker02}. In contrast, both Fang {\it et al.} and Bertazzi {\it et al.} calculated peak velocities of about $2.8\times10^7$ cm/s \cite{Fang19,Bertazzi09}, while a value of $\sim 2.6$-$2.7\times10^7$ cm/s is reported in Ref.~\cite{Wu05}. Fang {\it et al.} used a very similar DFT approach to that used here, with ONCV pseudopotentials and LDA XC functionals. The peak velocity and other differences are likely attributable to the use of a different pseudopotential-XC combination~\cite{Gaddemane20}.

For low fields (below $\sim$~50~kV/cm), our results follow very closely those of Refs.~\cite{Wu05,Yamakawa09,Fang19}, suggesting similar electron mobilities. From the slope at these low fields, we extract a mobility of $\sim$~2000~cm$^2$/(V$\cdot$s). Indeed, this agrees well with the reported value of 1950~cm$^2$/(V$\cdot$s) from Ref.~\cite{Fang19}. Both significantly exceed the experimentally measured value of 1300~cm$^2$/(V$\cdot$s) \cite{Webb01} (clearly seen by comparing the low-field slopes to that of Ref.~\cite{Barker02} in Fig.~\ref{fig:el_dv}). Fang {\it et al.} argue that this discrepancy is likely attributable to the lack of impurity and dislocation scattering in the theoretical models~\cite{Fang19}. To this point, in Fig.~\ref{fig:el_dv}, the slopes of Refs.~\cite{Bertazzi09,Albrecht98}, in which ionized impurity scattering is included, are closer to experiment. Additionally, it is likely that inaccuracies in the band structure play a significant role. For example, our calculated electron effective mass underestimates the measure value (see section \ref{sec:method.abinitio}), which one may predict will lead to a higher mobility. In a paper by Vitanov {\it et al.}, in which they employed a non-parabolic model using experimentally calibrated material parameters (including the effective masses), an electron mobility of $\sim$~1600~cm$^2$/(V$\cdot$s) is reported~\cite{Vitanov07}. As this paper does not take impurity or defect scattering into account, it is clear that improvements in the band structure cause appreciable improvement in the predicted mobility.

For the holes (Fig.~\ref{fig:el_h_E_dv}, right frame), we observe a much smaller low-field slope, and we extract a mobility of $\sim$~37~cm$^2$/(V$\cdot$s). This result is consistent with the relatively flat bands near the valence band maximum (Fig.~\ref{fig:band_structure}). While it is considerably lower than results of other first-principles calculations, which found a value of 52 cm$^2$/(V$\cdot$s) \cite{Ponce19}, it is in excellent agreement with Hall-effect measurements of 31 cm$^2$/(V$\cdot$s) \cite{Horita17}.

Following the peak, in all cases, the electron drift velocity decreases significantly and gradually reaches a steady-state saturation velocity. Chen {\it et al.} explain that saturation occurs as electrons fill the U and second $\Gamma$ valleys \cite{Chen08}. This work predicts a saturation velocity of approximately $1.1\times10^7$ cm/s, which slightly exceeds that of Ref.~\cite{Wu05}: $1.0\times10^7$ cm/s. Results from Refs. \cite{Fang19,Bertazzi09} are somewhat higher, settling down around $1.3\times10^7$ cm/s. The hole drift velocity saturates to $\sim 8.2\times10^6$ cm/s for a field strength of 1 MV/cm. As with the electrons, this is slightly lower than the value predicted by Bertazzi {\it et el.}, who calculated a saturation velocity of 9.4$\times10^6$ cm/s at the same field strength \cite{Bertazzi09}. Saturation for holes occurs as they fill the M and A valleys.

The kinetic energy-field curves for electrons and holes are shown in the left frame of Fig.~\ref{fig:el_h_E_dv}. We see that, initially, the electron energy remains nearly thermal until about 10 kV/cm, after which a gradual increase is observed. Between approximately 100-300 kV/cm, there is a sharper incline, which coincides with the peak electron drift velocity (large slope in the band structure). From 300-800 kV/cm, the slope of the energy-field curve gradually decreases as the average electron drift velocity saturates. These results agree well with those of Ref.~\cite{Fang19}. The characteristic ``S'' shape is also reported by Kolnik {\it et al.}~\cite{Kolnik95}, who used an empirical pseudopotential method to calculate the band structure. The major discrepancy between our work and that of Ref.~\cite{Kolnik95} is that electrons, in Ref.~\cite{Kolnik95}, remain near thermal energies up to much higher fields: up to about 50-100 kV/cm. This discrepancy is likely attributable to a difference in the electron effective mass (Ref.~\cite{Kolnik95}: 0.2~$m_e$, this work: 0.17~$m_e$). The larger effective mass reflects a smaller curvature of the band structure in the $\Gamma$-valley. Additionally, their method includes ionized impurity scattering, which, as we have already seen, decreases the mobility significantly. This likely increases the total scattering rate enough to prevent significant kinetic energy gains for fields below 50-100 kV/cm. 

\subsection{\label{sec:FBMC.therm}Hot Electron and Electron-Hole Pair Thermalization}
\begin{figure}[!]
\includegraphics[height=3.5in, width=3.5in]{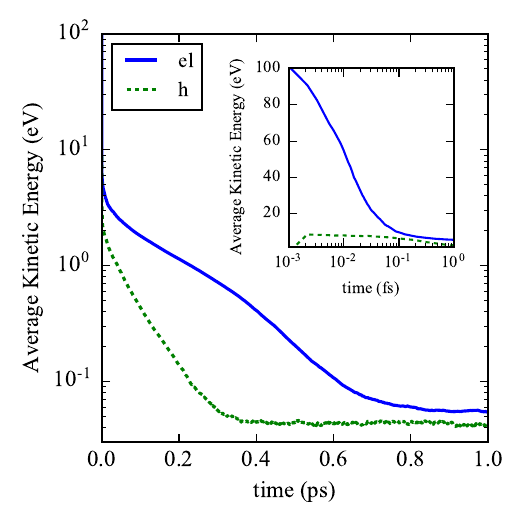}%
\caption{\label{fig:therm_semilogy}The time evolution of the injected hot-particle energy in wurtzite GaN at a temperature of 300 K with no applied field. The entire duration of the simulation is depicted in the larger frame, while only the first femtosecond is shown in the inset frame. The inset figure is included to observe more easily the rapid energy loss due to plasmon emission (and some impact ionization) at the beginning of the simulation.}
\end{figure}
\begin{figure*}
\includegraphics[height=5.5in, width=6.5in]{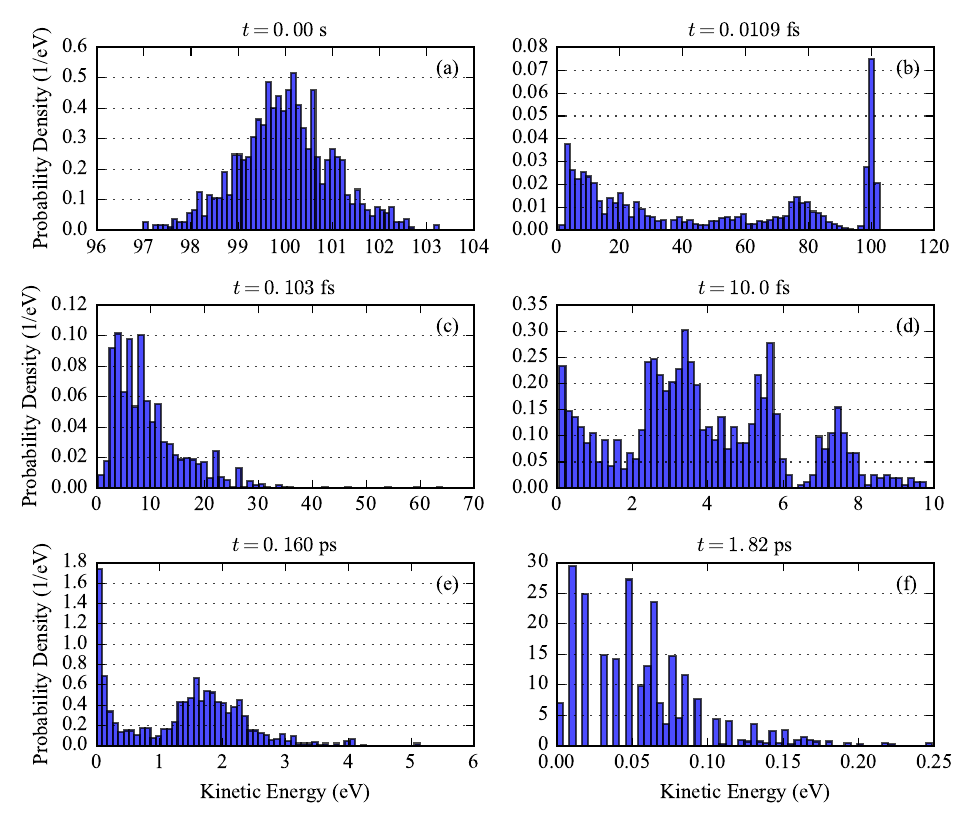}
\caption{\label{fig:E_distr}The time evolution of the energy distribution of injected hot electrons in wurtzite GaN at a temperature of 300 K with no applied field. The distribution functions have been normalized to the total number of electrons.}
\end{figure*}

For the hot-electron thermalization simulation, we begin with a bulk wurtzite GaN crystal at a temperature of 300 K with no applied field. In such a system, at equilibrium, the electrons are at energies near the thermal level ($\sim$~40~meV). The sudden appearance of hot electrons is simulated by “injecting” a set of 1000 electrons with energies under a Gaussian distribution centered at 100 eV, as described in section \ref{sec:FBMC.init_FF}.

In Fig.~\ref{fig:therm_semilogy}, we plot the average electron and hole energies as a function of time, including the initial hot electrons and the generated EHPs. The time required for full electron thermalization is approximately 1 ps. For generated holes, thermalization is complete in roughly half the time. The initial steps of the simulation are difficult to see in the larger plot, due to the x-axis scale, so an inset figure is included (upper right), with a log time-scale. In this inset figure, we see that the electron energy drops sharply from 100 to 10 eV in 0.1-0.2 fs. This rapid decline of the kinetic energy per particle is caused by the fast transfer of the initial kinetic energy to generated carriers via the very fast emission of plasmons at these energies, with an emission rate of the order of 10$^{16}$-10$^{17}$~s$^{-1}$ (Fig.~\ref{fig:ELR_valence band_DOS_JDOS}). 

The emitted plasmons tend to possess energies of $\sim$~21, 24, and 28~eV, due to corresponding peaks in the loss function (Fig.~\ref{fig:lor_fit}). Figure~\ref{fig:E_distr} shows several snapshots of the energy distribution at certain times throughout the simulation. Initially (Fig.~\ref{fig:E_distr} (a)), the distribution is Gaussian, as expected. At 0.0109~fs (Fig.~\ref{fig:E_distr} (b)), we see that many electrons have already experienced one or more plasmon emission events. Peaks at roughly 80, 60, and 40~eV indicate losses in multiples of the abovementioned plasmon peak energies. The large number of electrons with energies below 20 eV correspond to generated EHPs. After 0.103~fs, we see, in Fig.~\ref{fig:therm_semilogy}, a sudden decrease in the thermalization rate. The energy distribution at this time indicates that most electrons possess energies below the lowest plasmon energy (Fig.~\ref{fig:E_distr} (c)), causing plasmon emission to cease and impact ionization and phonon scattering to be the primary scattering mechanisms. In Fig.~\ref{fig:ELR_valence band_DOS_JDOS}, we see that with the dissipation of plasmon emission below $\sim$~20~eV, the scattering rate drops an order of magnitude to 10$^{15}$ s$^{-1}$, which causes this decrease. After 10.0~fs (Fig.~\ref{fig:E_distr} (d)), the tail of the energy distribution, on the high end, no longer exceeds 10 eV. All electrons in the simulation, therefore, experience only phonon scattering. 

Looking now at holes, we do not observe the same rapid energy decrease at the beginning of the simulation. Unless core states have been ionized by the original irradiation (a situation that we ignore here), hot holes do not possess sufficient energy to exceed the lowest plasmon frequency. Instead, holes tend to lose energy by a combination of impact ionization and phonon emission, leading to their much flatter slope in Fig.~\ref{fig:therm_semilogy} (inset figure; $t\le0.1$~fs). We note the increase in the average hole energy at the beginning of the simulation (inset figure; $t\le0.003$~fs). Initially, no holes are included, and this increase simply corresponds to hole buildup as EHPs are generated. In the later stages of the process, which we can see in the larger frame of Fig.~\ref{fig:therm_semilogy}, the slope of the hole thermalization is greater than that of the electrons. The greater slope can be explained by the fact that hole-phonon scattering rates below $\sim$~2.5~eV are roughly one order of magnitude larger than those of electrons (Fig.~\ref{fig:ph_rates}).

\begin{figure}[!]
\includegraphics[height=3in, width=3.5in]{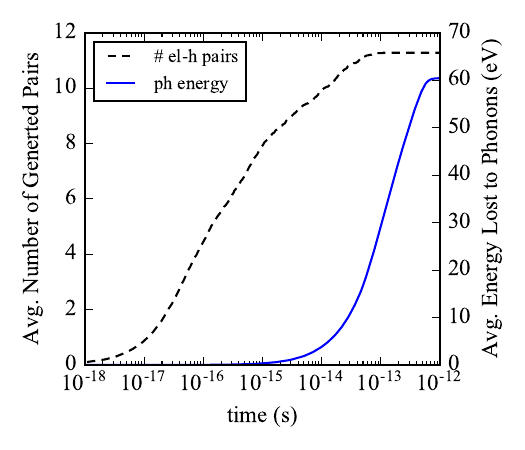}%
\caption{\label{fig:avg_pairs_avg_ph_E}Dashed line: The average number of pairs generated per hot electron as a function of time. Saturation occurring at 11.3 suggests an average generation of 11-12 pairs per hot electron. Solid Line: The average energy lost to phonons per hot electron as a function of time.}
\end{figure}

To observe where the energy goes in the thermalization process, the average number of generated pairs (per hot electron) is plotted along with the average energy lost to phonons as a function of time (Fig.~\ref{fig:avg_pairs_avg_ph_E}). We see that during the initial stages of the simulation ($t<1$ fs), most of the energy losses are to pair generation, as the energy lost to phonons is practically zero. At $\sim$~10$^{-14}$~s, phonon emission increases significantly, and the pair generation rate begins to flatten. These phonons will eventually decay, potentially resulting in temperature increases in the material. 

In this work, temperature effects are ignored, as we focus instead on the energy-loss processes in the ``10-100 eV gap''. As phonons do not occur early on, while most electrons are still in the ``10-100 eV gap'', it is not anticipated that temperature effects will significantly influence thermalization in this regime. Depending on the phonon density later in the process, temperature rises may be significant, affecting transport characteristics and device operation. Here we assume a low radiation dose and dose rate, resulting in a low hot electron density and, therefore, a low phonon density. This would lead to quick energy dissipation and removal from the system, and, therefore, to minimal temperature rises.

One may notice that $\sim$~60\% of the initial energy is lost to phonons by the end of the simulation. The other $\sim$~40\% can be accounted for by the ionization energy of each pair (number of pairs times the gap) plus some leftover kinetic energy of the electrons and holes. Eventually, this ionization energy will be lost to radiative (photons) and non-radiative (phonons) electron-hole recombination, and to Auger recombination. It is, however, presumed that the timescale for recombination will be significantly greater than a picosecond, and, therefore, it is not accounted for in this work. Indeed, the measurements of Jursenas {\it et al.} support this assumption, as the smallest recombination lifetimes are found to be of the order of a nanosecond \cite{jursenas_2005}.
\begin{figure*}
\includegraphics[height=5.5in, width=5.5in]{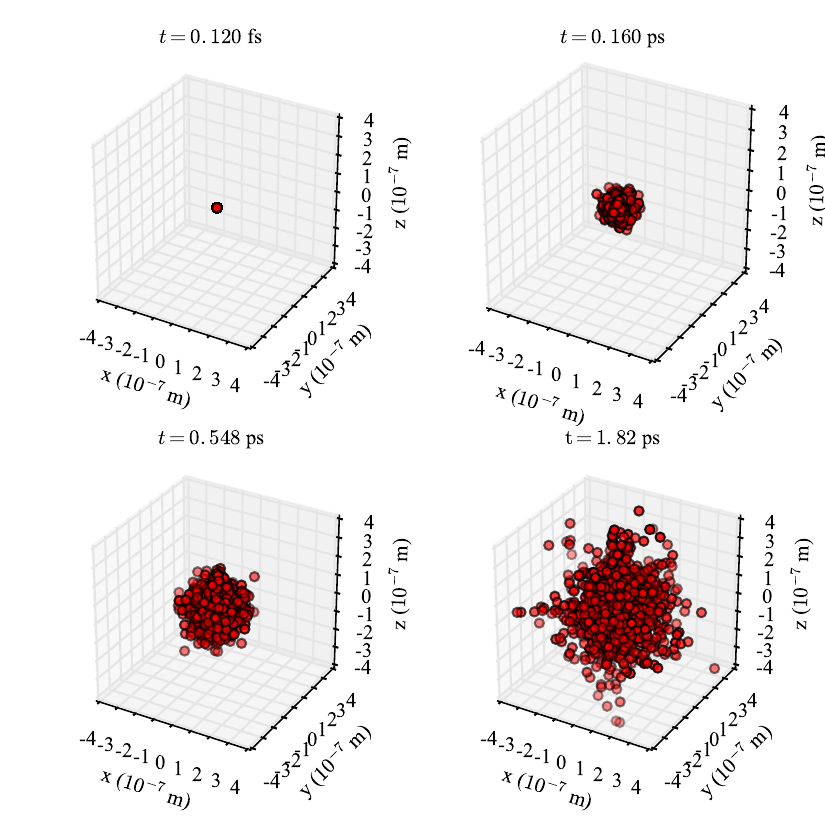}
\caption{\label{fig:el_pos_vs_t}The time evolution of the real-space position of the hot electrons. This simulation begins with all electrons at $(0,0,0)$ to observe how they travel and spread over time. It does not represent a real-life situation; it is merely an experiment.}
\end{figure*}

In addition to the energy distribution, we have also tracked the real-space positions of the simulated particles. This tracking allows us to analyze the real-space spread and average distance travelled during thermalization. We start the simulation with all electrons at the origin (Fig.~\ref{fig:el_pos_vs_t}). By the end (Fig.~\ref{fig:el_pos_vs_t}, bottom, right), the average distance travelled is on the order of 100~nm. As modern transistors have dimensions of approximately tens of nm, which continue to shrink, this result suggests that electrons generated by ionizing radiation will easily reach the boundaries of the device and may travel through several devices. This conclusion is in agreement with Weller {\it et al.}~\cite{weller_2004}, who emphasize that this must be true for electronic equilibrium as a condition for proper device simulation and testing.

Lastly, we calculate the average energy required to generate an electron-hole pair, which is commonly referred to as the mean ionization energy. This number is important for understanding and determining the amount of free charge created by ionizing radiation/particles in electronic devices, especially in binary-collision codes. Furthermore, it is a parameter used in determining performance characteristics of semiconductors in radiation detection. By simply tracking the energies of EHPs as they are created, we calculate an average creation energy of $\sim$~8.9~eV (with each hot electron generating 11-12 pairs, throughout the process). It is generally accepted that the mean ionization energy in a semiconductor is approximately equal to three times the bandgap ($3\times3.4=10.2$ eV, for GaN). On first glance, our result is reasonable, as it is within $\sim$~14\% of this guideline. A number of analytic approximations are also available to calculate this energy~\cite{shockley_1961,pantelides_2022,Klein68}. Using the empirical expression reported by Klein \cite{Klein68},
\begin{equation}
E_{\rm i} = 2.8 E_{\rm g} + 0.6 \ {\rm eV},
\end{equation}
we calculate a value of 10.12 eV. A lower value of 9.59~eV has been found via electron-beam induced current measurements \cite{Yakimov21}. Even better (if not identical) agreement is observed with Ref.~\cite{Sellin06}, in which a value of 8.9 eV is reported. Overall, we conclude that the calculated average creation energy is reasonable. 

\section{\label{sec:conclusions}Conclusions}
We have presented a first-principles study of hot-electron and EHP thermalization in wurtzite GaN to close the ``10-100 eV gap''. We have developed a FBMC code in which we have included plasmon emission and impact ionization, calculated using the full dynamic dielectric function, for accurate simulation of the thermalization processes in the intermediate energy range ($\sim$~10-100~eV). We also include phonon scattering for all valence bands (including the {\it d} bands) and for conduction bands up to a kinetic energy of $\sim$~25 eV. We have found that, in agreement with Pines~\cite{Pines_1956}, plasmon-mediated processes dominate at high energy (during the initial 0.1 fs), impact ionization at intermediate energies (from 0.1 to $\sim$~10 fs), and that phonons control the later stages of the thermalization (from 10 fs to full thermalization). 

In addition to studying the time-scale, we have also investigated the length-scale (diffusion of hot carriers) and found that hot carriers travel an average distance of $\sim$~100 nm. As others have found~\cite{weller_2004,Reed15}, this distance easily exceeds the typical dimensions of modern electronic devices, consistent with the requirement to establish secondary electronic equilibrium during device simulation and testing. It is important to know how fast and how far these carriers decay as they may create defects in the crystal, cause unexpected and unwanted transients and shorts in FETs, and cause catastrophic failure of the device~\cite{Reed15}.

We have also calculated an average EHP creation energy of $\sim$~8.9 eV/pair which agrees well with reported data. This number, also known as the mean ionization energy, is a critical parameter for the binary-collision codes of the nuclear/particle physics community. It also gives an idea of the amount of excess charge to expect, and, therefore, what kind of damage to predict. 

Overall, this work provides the understanding, methods, and framework to study theoretically the full thermalization of charge carriers generated from a realistic streak of an ionizing particle in an electronic device.

\section{\label{sec:acknowledgements}Acknowledgements}
We would like to thank A. Edwards and K. Goretta for organizing this research project and leading our many discussions. This work has been supported by the Air Force Office of Scientific Research: Award FA9550-21-0461, and through the Center of Excellence in Radiation Effects: Award FA9550-22-1-0012.
\bibliography{don_paper}

\begin{thebibliography}{101}
\bibitem [{foo()}]{footnote_rejtech}%
  \BibitemOpen
  \href@noop {} {}\bibinfo {note} {The rejection technique is a standard
  technique to generate a variable $x$ in ($x_{min}$,$x_{max}$) with a given
  probability distribution $P(x)$ with maximum $P_{max}$ in
  ($x_{min}$,$x_{max}$): Two random numbers, $r_1$ and $r_2$, are generated
  with uniform distribution in (0,1). The value $x_{min}+r_1(x_{max} -
  x_{min})$ is selected only if $r_2P_{max}<P[x_{min}+r_1(x_{max}-x_{min})]$,
  otherwise the process is repeated (that is, the value $x_{min}+r_1(x_{max} -
  x_{min})$.}\BibitemShut {Stop}%
\bibitem [{syn()}]{sync_ens}%
  \BibitemOpen
  \href@noop {} {}\bibinfo {note} {Monte Carlo simulations of particle
  transport can be performed considering a single particle at the time or,
  instead, considering an ensemble of many particles that are tracked
  synchronously. This permits the study of time dependent
  phenomena.}\BibitemShut {Stop}%
\end{thebibliography}

\end{document}